\documentclass{aa}
\pdfoutput=1
\usepackage{graphicx}
\usepackage{txfonts}
\usepackage{mathrsfs}
\usepackage{multirow}
\usepackage{ulem}
\usepackage{amsmath}
\usepackage{xcolor}
\usepackage{orcidlink}
\usepackage{placeins}
\usepackage{hyperref}
\hypersetup{colorlinks=true,linkcolor=blue, filecolor=magenta,citecolor=blue,urlcolor=cyan}
\newcommand{\xmm}{{\textit{XMM-Newton}}\/}

\newcommand{\swift}{{\textit{Swift}}\/}

\newcommand\T{\rule{0pt}{2.9ex}}                                         
\newcommand\B{\rule[-1.5ex]{0pt}{0pt}} 

\begin{document}

\title{X-ray eclipse tomography: Resolving the extent of X-ray-emitting region(s) in the Seyfert galaxy ESO~362-G18}

\titlerunning{Resolving the extent of X-ray-emitting region(s) in ESO~362-G18}

\authorrunning{Rueda-Nieto et al.}

   \author{
   Laura Rueda-Nieto  
   \inst{1,2} 
   \orcidlink{0009-0008-5759-4299}
   \and
   Beatriz Ag\'is-Gonz\'alez
   \inst{3} 
   \orcidlink{0000-0001-7702-8931}
   \and
   Giovanni Miniutti 
   \inst{4} 
   \orcidlink{0000-0003-0707-4531}
   }
   
   \institute{
   Universidad de Alcal\'a, Plaza San Diego s/n, E-28801 Alcalá de Henares, Madrid, Spain \\
   \email{laura.ruedan@edu.uah.es}
    \and 
    Universidad Complutense de Madrid, Av. Complutense s/n, E-28040 Madrid, Spain 
    \and
    Institute of Astrophysics, Foundation for Research and Technology-Hellas, GR-70013 Heraklion, Greece
    \and
    Centro de Astrobiolog\'ia, CSIC-INTA, Camino Bajo del Castillo s/n, E-28692 Villanueva de la Ca\~nada, Madrid, Spain 
    }

\date{Received  / Accepted  }
\date{}

\abstract{
X-ray spectral variability in active galactic nuclei (AGNs) can be either intrinsic or caused by external processes, such as variable absorption. Several transient occultation events from individual clouds or X-ray eclipses have been identified in a number of AGNs in the past two decades. We report results from the analysis of two \swift\ monitoring campaigns of the Seyfert galaxy ESO~362-G18, a changing-look AGN that showed transitions between optical spectral types 1.5 and 1.9 in the past, most likely induced by variable absorption. We identified two X-ray eclipses during the first \swift\ campaign, one of which we were able to follow almost entirely from ingress to egress. We used time-resolved spectroscopy to follow the evolution of the X-ray spectrum with time in order to derive the properties of the absorbing cloud and the size of the main X-ray-emitting region. The cloud is located towards the innermost dust sublimation zone, intermediate between the dust-free broad-line region and the dusty torus. The X-ray-emitting region is confined within $\sim 35$ gravitational radii ($R_g$) from the centre. A deeper analysis indicated that the eclipsing cloud likely comprises a denser core and a more tenuous atmosphere. We were also able to estimate the size of the soft excess and hot corona X-ray-emitting regions separately. These are two of the most relevant continuum components in the X-ray spectra of unobscured AGNs. Our analysis suggests that the soft-excess-emitting region is $\sim 50$\% more extended than the hot corona one. However, it appears to co-exist with and is not replaced by the hot corona one in the innermost $\sim 30~R_{\rm g}$. Our results highlight the potential of X-ray eclipse tomography for providing a clearer view of the innermost accretion flow geometry and of the AGN surroundings out to the torus scale.
}

\maketitle


\section{Introduction}
\label{sec:intro}

Active galactic nuclei (AGNs), powered by accretion onto supermassive black holes (SMBHs), are among the most energetic sources in the Universe \citep{Shakura1973}. They emit radiation across the entire electromagnetic spectrum, including ubiquitous X-ray emission \citep{Elvis1994,Lusso2016}, and are variable on a wide range of timescales \citep{Lawrence1987, Peterson1993,McHardy2006}. In X-rays, the typical spectrum of an unobscured (type I) AGN consists of several emission components. These include (1) a hard power-law continuum produced via inverse-Compton scattering in a hot optically thin corona \citep{Sunyaev1980,Haardt1993,Haardt1994}; (2) a soft X-ray excess \citep{Laor1997, Porquet2004}, whose origin remains debated and may arise from a warm optically thick corona \citep{Petrucci2018}, from relativistic reflection off the inner accretion disc \citep{Crummy2006}, or from a combination of both processes \citep[e.g.][]{Xiang2022, Ballantyne2024}; and (3) X-ray reflection features \citep{George1991,Matt1991} originating from distant material and the accretion disc itself. Absorption is also present, often due to partially ionised outflowing gas along the line of sight \citep[e.g.][]{Blustin2005,Tombesi2013, Laha2014_wax1}. The interplay between the emission components, which can vary with different amplitudes and on different timescales \citep[e.g.][]{Arevalo2006}, can lead to intrinsic X-ray spectral variability. Spectral variability can also be extrinsic, caused by changes in the properties of the absorbing material, such as clouds or inhomogeneous winds crossing the line of sight \citep[e.g.][]{Risaliti2002,Elvis2004,Puccetti2007,Bianchi2009}. The study of these events provides a powerful tool for probing the inner circumnuclear environment, revealing the physical properties, geometry, and dynamics of gas and dust surrounding the SMBH and the physical extent of the emitting region.

The AGN X-ray spectral variability caused by absorption is observed over a wide range of timescales, from hours to years. This indicates that the obscuring material is distributed at different distances from the nucleus. Short-duration eclipses have revealed compact dust-free absorbers orbiting within the dust-sublimation radius and with overall properties similar to those of broad-line region (BLR) line-emitting clouds \citep[e.g.][]{Risaliti2005,Risaliti2007,Risaliti2009,Sanfrutos2013,Gallo2021}, while longer-duration events have been linked to clouds at larger distances, in the torus or intermediate regions \citep[e.g.][]{Lamer2003, Miniutti2014}. Studies of X-ray absorption variability in statistically significant AGN samples further support the clumpy nature of the circumnuclear medium \citep[see e.g.][]{Markowitz2014,Torricelli-Ciamponi2014,Lian2025}. These results provide direct evidence of a circumnuclear medium  composed of discrete structures, likely embedded in more homogenous media \citep{Maiolino2010comet,Sanfrutos2016}, with important implications for unified models of AGNs \citep{Antonucci1993_unification}.

We study the AGN ESO 362-G18 (also known as MCG 05-13-17), a Seyfert galaxy \citep{Bennert2006} at $z=0.012$ that has been extensively observed across multiple wavelengths. Integral field spectroscopy reveals a disturbed galactic structure, possibly due to a minor merger, as well as ionised gas in rotation and large-scale outflows \citep{Humire2018}. Historical optical spectra showed transitions in spectral type, and the target can be classified as type 1.9 (2003, 2006) or 1.5 (2004, 2016), the fastest being a 1.9 $\rightarrow$ 1.5 transition taking place in less than 20 months \citep{Bennert2006,Parisi2009,Agis2018}. The lack of high-amplitude intrinsic variability in historical data strongly suggests that the observed transitions are due to variable intervening absorption along the line of sight towards the BLR, making ESO~362-G18 a strong absorption-induced changing-look AGN candidate \citep{Agis2017PhDThesis,Agis2018}. 

In X-rays, the spectrum of ESO~362-G18 is characterised by a standard hard X-ray continuum, soft excess, and X-ray reflection from distant material and likely the inner disc as well \citep{Agis2014_ESO362,Xu2021_ESO362}. The origin of the soft excess, as in many AGNs, remains uncertain, with a warm corona and relativistic disc reflection being plausible contributors \citep{Xu2021_ESO362,Zhong2022}. \citet{Agis2014_ESO362} analysed a series of X-ray observations of ESO~362-G18 by X-ray Multi-mirror Mission - Newton (\xmm), Chandra X-ray Observatory (\textit{Chandra}), Neil Gehrels Swift Observatory (\swift), and Suzaku X-ray satellite (\textit{Suzaku}) over the course of a few years and reported on spectral variability associated with a large increase in X-ray absorption during one of the \xmm\ observations that followed, by about two months, a basically unabsorbed \swift\ observation. They interpreted the data as the signature of a transient occultation event, or X-ray eclipse, from a cloud located towards the innermost region of the dusty and clumpy torus. From the observed eclipse, they were able to estimate that the X-ray-emitting region is confined within $\simeq 48~R_g$ ($R_g = GM_{\rm BH}/~c^2$) from the central SMBH. The absorption event suggests a line of sight that grazes the clumpy torus and thus enhances the probability of transient eclipses. This agrees with the changing-look properties of the AGN. \citet{Agis2014_ESO362} used X-ray relativistic reflection spectroscopy to estimate a line-of-sight inclination of $\simeq 53^\circ$, which is fully consistent with such a hypothesis. The SMBH mass in ESO~362-G18 is estimated to be  $M_{\rm BH}\simeq 4.5 \times10^7\,M_\odot$, and ESO~362-G18 has a typical bolometric luminosity of $L_{\rm bol}\sim1.3\times10^{44}\,{\rm erg/s}$, corresponding to an Eddington ratio of $\lambda_{\rm Edd}\simeq0.02$ \citep{Agis2014_ESO362}.

The goal of this work is to study the X-ray spectral variability of ESO 362–G18 using a series of observations obtained with a cadence of only a few days with the \textit{Neil Gehrels Swift} Observatory (\swift) over two epochs. We note that \citet{Lian2025} performed a systematic search for X-ray eclipses in AGNs in the \swift\ data archive. Only campaigns comprising at least 90 \swift\ observations or that were associated with continuous monitoring with at least 50 pointings were considered in their work. ESO~362-G18 was monitored by \swift\ on two occasions with campaigns comprising 36 and 35 observations, and it was therefore excluded by their selection criteria. 

We focus on the first monitoring campaign by \swift, and the analysis of the second epoch is briefly presented in the appendix. The paper is organised as follows. Section~\ref{sec:obs} discusses the Swift observations and data reduction. In Sect.~\ref{sec:specvar} we describe the X-ray spectral variability of ESO~362-G18 during the first monitoring campaign, while Sect.~\ref{sec:spec_var_epoch1} presents its time-resolved spectral analysis. The results from the spectral analysis are used to estimate the properties of the X-ray-emitting regions and of the ambient medium in ESO~362-G18 in Sects.~\ref{sec:properties} and \ref{sec:onestep}. We discuss our results in Sect.~\ref{sec:discussion}. Appendix~\ref{app:Xrays_epoch1} presents the details of the X-ray spectral analysis during the first \swift\ campaign and the analysis of X-ray data from the second campaign, while Appendix~\ref{app:UVOT} describes the simultaneous \swift\ UV-Optical Telescope (UVOT) observations in the optical and UV during both epochs. 

\section{Observations and data analysis}
\label{sec:obs}

ESO~362-G18 has been observed in X-rays several times and with different observatories in the past \citep[see for example][]{Agis2014_ESO362,Xu2021_ESO362}. We focused on \swift\ monitoring observations during two different epochs, the first between MJD 55515 and 55584 (epoch~1, comprising 36 observations), and the second between MJD 59896 and 59959 (epoch~2, 35 observations with a $\sim 10$~d gap in between). As mentioned, we focused on the first epoch here, and the analysis of data from the second campaign will be presented in the appendix.

The monitoring during epoch~1 was conducted to follow the optical and UV evolution of the Type~IIb supernova SN2010jr, which was first detected on 2010 November 12 (MJD 55510) about 15\arcsec\ from the nucleus \citep{SN2010jr_1,Pritchard2014}. We verified that even during the first monitoring observation when SN2010jr was at its brightest, its X–ray flux contribution was negligible, with a broad-band X-ray flux $\geq 20$ times fainter than the nuclear X-ray emission and dimming quickly with time \citep{SN2010jr_2}. 

We made use of observations performed with the \swift\ X-ray Telescope (XRT) in photon-counting (PC) mode that were analysed following the procedure outlined in \citet{Evans2007,Evans2009}, which uses fully calibrated data and corrects for effects such as pile-up and the bad columns on the CCD, to obtain count rates and spectral products on an observation-by-observation basis. Specifically, light curves  and spectra were obtained (and combined when necessary) using the online XRT product generator tool\footnote{\url{https://www.swift.ac.uk/user\_objects}} after checking that products were indistinguishable from those obtained through manual extraction. 

All X-ray spectra we used were grouped to a minimum of one count per bin, and we used the \texttt{xspec} \citep{Arnaud1996xspec} implementation of the Cash statistics \citep{Cash1979} for parameter estimation. Uncertainties were estimated via Monte Carlo Markov Chain simulations based on the Goodman–Weare sampler \citep{GoodmanWeare_MCMC2010} as implemented into \texttt{xspec}. We used $100$ walkers and $1.6\times 10^6$ steps, rejecting the first $6\times 10^5$ ones as burn-in. We report uncertainties corresponding to $90$\% credible intervals. 

For the model comparison, we applied the Akaike Information Criterion \citep[AIC;][]{Akaike1974_AIC} corrected for finite sample size \citep{Sugiura1978}, which can be used as an approximate method of model selection based on the likelihood ratio and which penalises model complexity. Models with the lower AIC should be preferred and a difference $\Delta {\rm AIC} = {\rm AIC}_{1} - {\rm AIC}_{2} > 10$ between two competing models is generally considered as an indication of strong preference for the second model over the first \citep[see e.g.][]{Emmanoulopoulos2016}.  

The \swift\ UV/optical telescope (UVOT) was operated simultaneously to the XRT taking data in all of the available filters from V (centred at $\sim 5468$~\AA) to UVW2 ($\sim 1928$~\AA). Since we were only interested here in the central AGN, during epoch~1 we extracted optical/UV products taking care of excluding SN2010jr from the source and background regions using a circular region of 5\arcsec\ radius for the source, and an annulus with inner (outer) radii of 6\arcsec\ (12\arcsec). Count rates and flux densities in all filters were extracted via  the dedicated \texttt{uvotsource} task using the option \texttt{ssstype=low} to identify and exclude observations in which the source lied onto a low-sensitivity area of the UVOT detector. Results were unaffected when the more conservative cut \texttt{ssstype=mid} was used instead. When necessary, the photometric data were combined with background and responses for joint spectral analysis with the X-ray data from XRT using the \texttt{uvot2pha} task.

\section{The \swift\ campaign}
\label{sec:specvar}

\begin{figure}[t]
\centering 
\includegraphics[width=0.95\columnwidth]{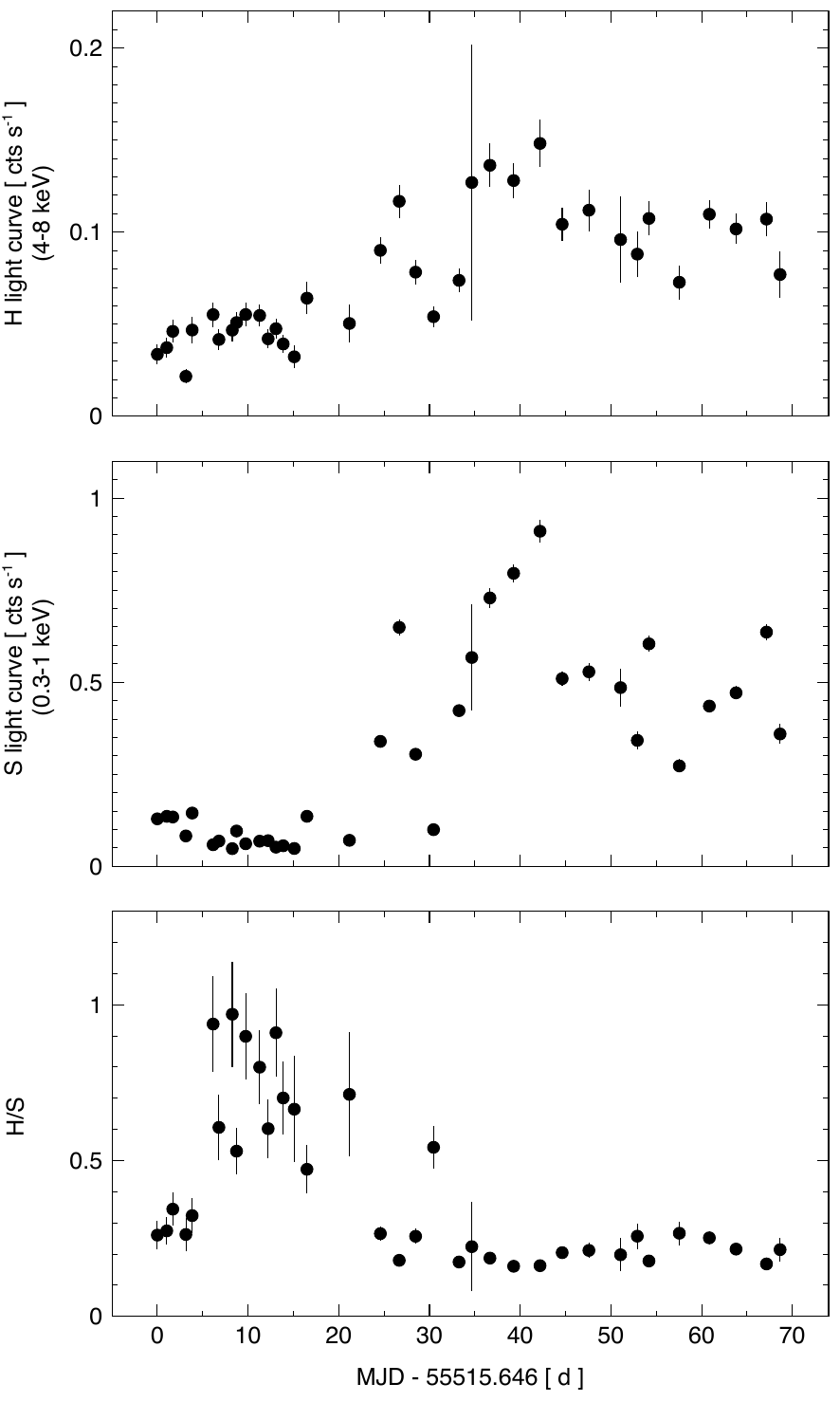}
\caption{\textit{Top and middle}: H 4-8~keV and S 0.3-1~keV light curves, respectively. \textit{Bottom}: Corresponding H/S.}
\label{fig:lc}
\end{figure}

We focused on \swift\ observations during epoch~1, while the analysis of \swift\ data from epoch~2 is reported in Appendix~\ref{app:X-rays_epoch2}. The hard (H) 4-8~keV light curve from the first \swift\ campaign is shown in the upper panel of Fig.~\ref{fig:lc}. The source is variable with a minimum to maximum variation of a factor of $\sim 7$. To search for any spectral variability, we considered also a soft band (S) defined in the 0.3-1~keV range, and we computed the hard-to-soft ratio (H/S). The S and H/S light curves are shown in the middle and lower panels of Fig.~\ref{fig:lc} respectively. One time interval of particularly high H/S stands out at the beginning of the campaign. 

\begin{figure}
\centering 
\includegraphics[width=0.95\columnwidth]{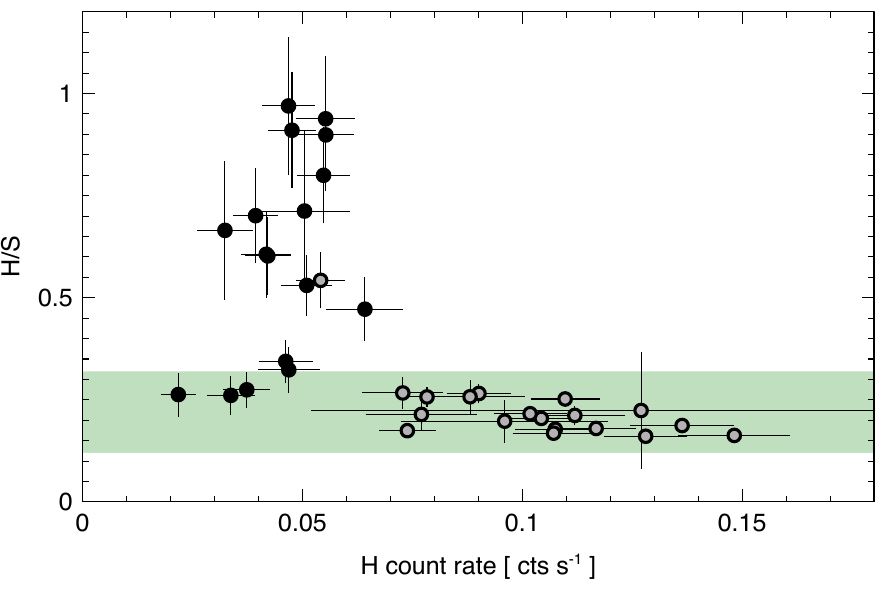}
\caption{Hard-to-soft ratio as a function of hard X-ray count rate (H). Black (grey) circles represent data from the initial $\sim 23$~d (latest $\sim 46$~d) into the campaign. The shaded area represents H/S~$=0.22 \pm 0.10$, which describes the approximately constant spectral shape during the second part of the campaign.}
\label{fig:HS_1}
\end{figure}

The H/S ratio as a function of H count rate is shown in Fig.~\ref{fig:HS_1} where we have separated data points from the initial $\sim 23$~d of the campaign (black symbols) from the later ones (light grey). The excursion into high H/S is confined within the first part of the campaign, while H/S is consistent with being constant ($H/S \simeq 0.22$) thereafter (shaded area in Fig.~\ref{fig:HS_1}). No correlation is seen between spectral shape (H/S) and hard X-ray count rate (H) despite an overall variation by about a factor of $\sim 7$. During the second part of the campaign, H/S remains approximately constant despite a variation by a factor of $2$-$3$. This suggests that any intrinsic spectral variability is negligible and that the H/S variation is entirely due to an extrinsic phenomenon occurring at the beginning of the monitoring campaign only. 

\section{X-ray spectral variability}
\label{sec:spec_var_epoch1}

\begin{figure}
\centering 
\includegraphics[width=0.95\columnwidth]{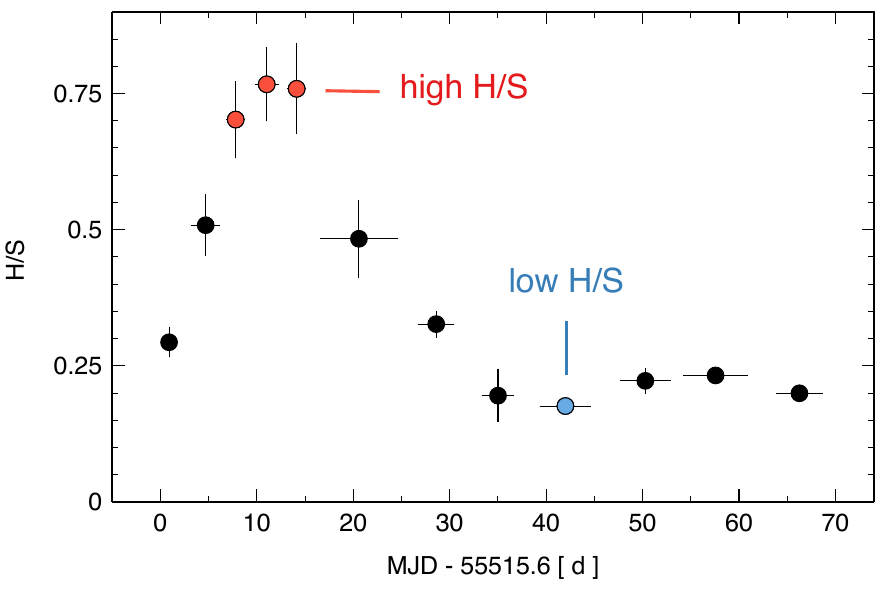}
\caption{Hard-to-soft ratio light curve re-binned by a factor of 3. We highlight data points associated with high and low H/S that were used to extract the low and high H/S X-ray spectra shown in the upper panel of Fig.~\ref{fig:HSspec}.}
\label{fig:epoch1_rebinned}
\end{figure}

A re-binned version (by a factor of 3) of the H/S light curve is shown in Fig.~\ref{fig:epoch1_rebinned}. The H/S light curve is characterised by a smooth bell-like feature peaking at $\simeq 12$~d into the campaign and lasting $\simeq 40$~d. A further, lower-amplitude maximum of H/S is also seen at $\simeq 56$~d. To study the origin of the observed spectral variability, we selected two time-intervals corresponding to high and low H/S respectively (highlighted in Fig.~\ref{fig:epoch1_rebinned}), and we extracted the corresponding X-ray spectra. The high H/S spectrum was extracted by combining 9 observations (ObsID 00031868007-00031868015), while the low H/S one was obtained by stacking 3 of them (ObsID 00031868025-00031868027). The unfolded X-ray spectra are shown in the upper panel of Fig.~\ref{fig:HSspec} using the same colour scheme as in Fig.~\ref{fig:epoch1_rebinned} together with their respective best-fitting models, that are described below. The high H/S spectrum differs from the low H/S by a lower intrinsic X-ray flux (as indicated by the hardest data points) and, most importantly, by being significantly more absorbed below $\sim 6$~keV \citep[for similar behaviour in the Seyfert galaxy ESO~323-G77, see e.g. Fig.~1 in][]{Miniutti2014}. 

\subsection{Analysis of the low and high H/S spectra}
\label{sec:low_high_HS}

A detailed description of the spectral analysis is given in Appendix~\ref{app:Xrays_epoch1} (see, in particular, Appendix~\ref{app:low_high_HSspec}). We briefly discuss the resulting best-fitting model. We adopted a standard Seyfert~1 X-ray spectral model comprising a soft X-ray excess described by the {\texttt{comptt}} model in {\texttt{xspec}} \citep{Titarchuk1994_comptt}, a hard Comptonisation component \citep[\texttt{nthcomp}:][]{Zdziarski1996_nthcomp,Zycki1999_nthcomp}, and a Gaussian emission line representing Fe~K$\alpha$ emission with fixed rest-frame energy at $6.4$~keV and width free to vary. The overall model was absorbed by the Galactic column density fixed at $1.35\times 10^{20}$~cm$^{-2}$ \citep{HI4PI2016} using the \texttt{tbabs} model with cross sections and abundances from \citet{Wilms2000_abund}.

We searched for a solution describing the two spectra simultaneously with the least number of variable parameters, and we encountered an excellent description of the data using a model with the following properties: (i) the intrinsic flux variability occurs at fixed spectral shape (consistent with the green shaded area in Fig.~\ref{fig:HS_1}), and (ii) the observed spectral variability is entirely associated with variable absorption. The variable absorption was implemented by allowing the covering fraction of an ionised absorber  with constant column density and ionisation to vary between the two spectra using the \texttt{zxipcf} model in \texttt{xspec} \citep{Reeves2008_zxipcf}. In {\texttt{xspec}} notation, each spectrum is then described by 
\begin{align} 
\texttt{tbabs}\times\texttt{zxipcf}\times\left(\texttt{comptt}+\texttt{nthcomp}+\texttt{zgaus}\right), \label{eq:specmodel}
\end{align}
where all parameters are the same for the low and high H/S spectra, except for the covering fraction of the ionised absorber. Any flux variability, assumed to occur at constant spectral shape (as indicated by the shaded area in Fig.~\ref{fig:HS_1}), is accounted for by an overall normalisation constant that is allowed to vary between the two spectra.

\begin{figure}
\centering 
\includegraphics[width=0.95\columnwidth]{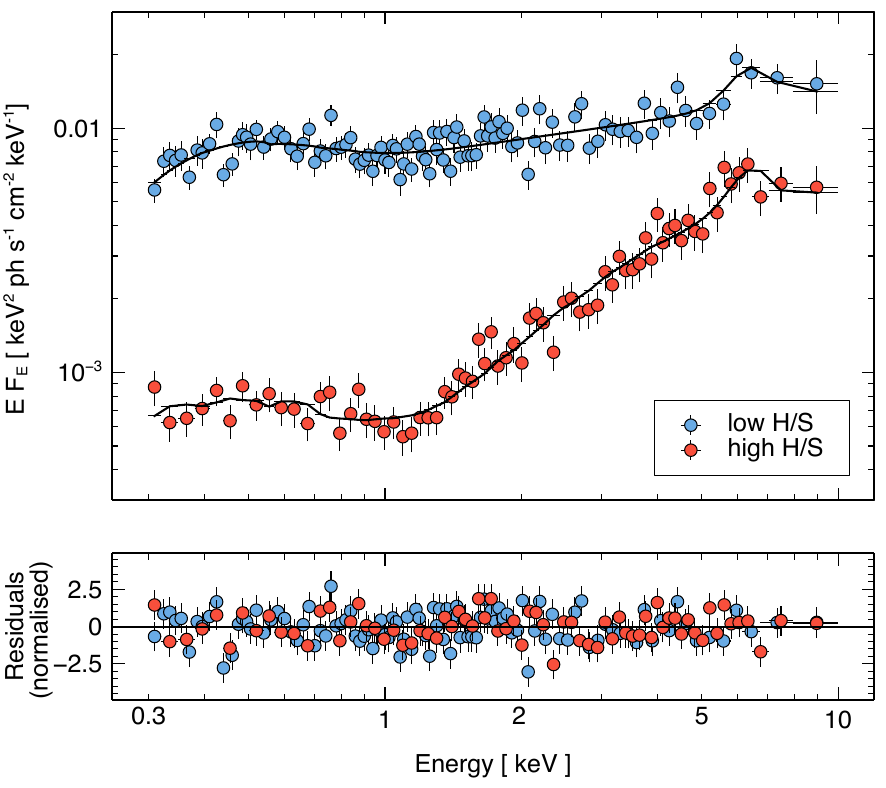}
\caption{\textit{Top}: Unfolded low (blue) and high (red) H/S X-ray spectra accumulated during the time-intervals highlighted in Fig.~\ref{fig:epoch1_rebinned} shown together with their respective best-fitting models. \textit{Bottom}: Resulting residuals normalised by the uncertainties. The data were re-binned for visual clarity.}
\label{fig:HSspec}
\end{figure}

This simple model, in which the only parameters that are allowed to vary independently are the intrinsic X-ray flux and the covering fraction $C_{\rm f}$ of the ionised absorber, resulted in an excellent description of the data with $C=1206$ for $1318$ degrees of freedom. The best-fitting models are shown as solid lines in the upper panel of Fig.~\ref{fig:HSspec} and the corresponding residuals, normalised by the uncertainties, are shown in the lower panel. The soft excess can be described by a warm optically thick corona with (common) temperature $kT_{\rm e} = 110\pm 10$~eV and poorly constrained optical depth $\tau \gtrsim 25$. The (common) slope of the hard Comptonisation component is $\Gamma = 1.75\pm 0.06$. A broad ($\sigma \simeq 300$-$700$~eV) Fe~K line at $6.4$~keV is detected and forcing the line to be unresolved resulted in a worse fit by $\Delta C = +11$ for 1 degrees of freedom. The quality of the short-exposure XRT spectra does not allow us to study the Fe line profile in detail nor to confidently claim the detection of a relativistically broadened Fe line. However, the line is resolved which is consistent with results by \citet{Agis2014_ESO362} who have analysed much higher quality data of ESO~362-G18, suggesting that a relativistic reflection component is present.

The only important difference between the low and high H/S spectra is associated with the covering fraction $C_{\rm f}$ of the ionised absorber. The (common) column density and ionisation are $N_{\rm H} = (5.0 \pm 1.1)\times 10^{22}$~cm$^{-2}$ and $\log \xi = 0.7\pm 0.4$, while $C_{\rm f} \leq 0.12$ for the low H/S spectrum, and $C_{\rm f} = 0.81 \pm 0.03$ for the high H/S one. As mentioned, the ionised absorber $C_{\rm f}$ accounts for all of the spectral variability. 

\subsection{Time-resolved spectroscopy}
\label{sec:time_resolved}

The smooth shape of the H/S light curve in Fig.~\ref{fig:epoch1_rebinned}, coupled with results from the low and high H/S spectral analysis, suggests that the variable H/S during the campaign is due occultation events (or eclipses) progressively covering and uncovering the X-ray-emitting region with maximum coverage around $\simeq 12$~d, and possibly also $\simeq 56$~d. We then considered a time-resolved X-ray spectral analysis considering the 12 X-ray spectra corresponding to the data points shown in Fig.~\ref{fig:epoch1_rebinned} which are constructed by stacking three consecutive of the original \swift/XRT observations to increase the signal-to-noise ratio. 

\begin{figure}[t]
\centering 
\includegraphics[width=0.95\columnwidth]{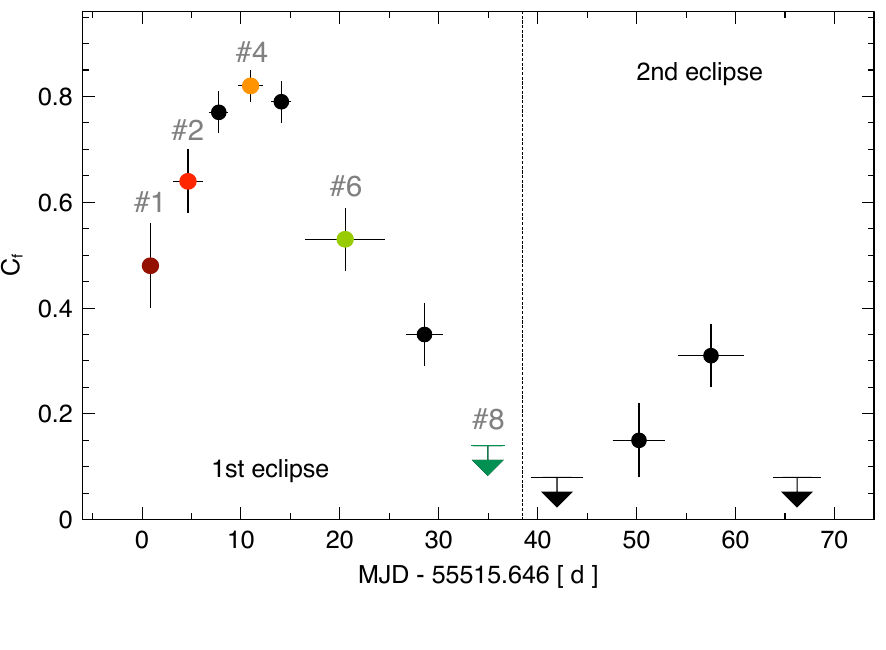}
\includegraphics[width=0.95\columnwidth]{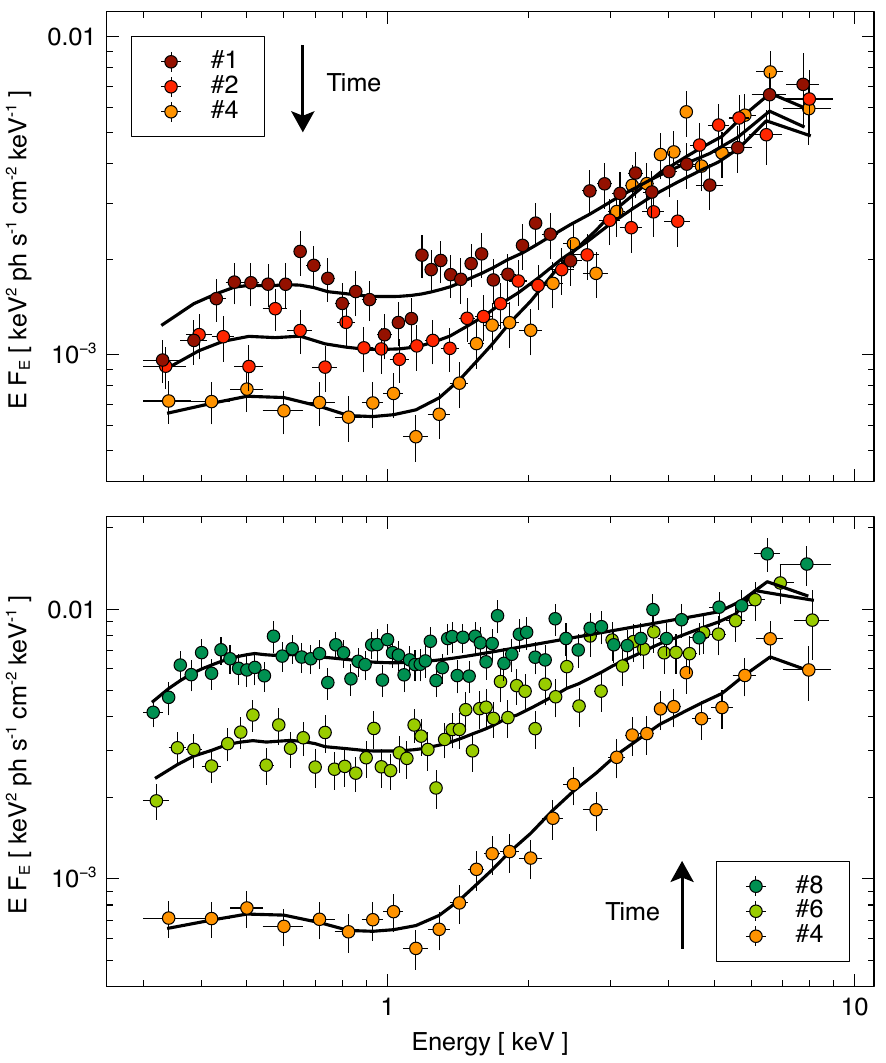}
\caption{{\it Top:} Evolution of the covering fraction from the time-resolved spectroscopic analysis using model~1 (Table~\ref{tab:table4}). The data points corresponding to the spectra shown in the lower panels are numbered and colour-coded. {\it Middle and bottom:} Selection of X-ray spectra from the time-resolved spectral analysis. Spectrum 4, corresponding to the $C_{\rm f}$ peak, is repeated in both panels. All X-ray spectra were re-binned for visual clarity.}
\label{fig:CF_spec}
\end{figure}

We adopted the same model described above (Eq.~\ref{eq:specmodel}) and fitted the 12 X-ray spectra simultaneously, allowing only an overall normalisation (governing the intrinsic X-ray flux) and the covering fraction $C_{\rm f}$ of the ionised absorber to vary independently. Details on the spectral analysis are reported in Appendix~\ref{app:time-resolved-spec}. The model provided a good description of the data with $C=5829$ for $6525$ degrees of freedom. The resulting covering fraction evolution is shown in the upper panel of Fig.~\ref{fig:CF_spec} and exhibits two well-separated maxima, corresponding to two distinct X-ray eclipses. The first eclipse encompasses the first eight data points (initial $35$-$40$~d), while the second, less pronounced eclipse is defined by the last four. 

Given that the two eclipses are likely from different structures - or clouds - we considered a further model in which the ionised absorber column density and ionisation were allowed to be different between the two events (but the same during each eclipse). We measured $N_{\rm H} = (4.5\pm 1.0) \times 10^{22}$~cm$^{-2}$ and $\log\xi = 0.65\pm 0.35$ during the first eclipse (first eight spectra), and $N_{\rm H} = (0.7\pm 0.5) \times 10^{22}$~cm$^{-2}$ with a poorly constrained  $\log\xi \leq 0.8$ during the second (last four spectra). The fit statistics was $C=5819$ for $6523$, and although the improvement cannot be considered as highly significant ($\Delta {\rm AIC} = 5.96$), we retained this latter model as the best-fitting one because it is most likely that two different absorbing structures have different $N_{\rm H}$ and $\log\xi$. The best-fitting parameters for this model, hereafter Model~1, are reported in Table~\ref{tab:table4}. Letting column density and/or ionisation free to vary independently between spectra for each eclipse did not improve the fitting statistics. The upper panel of Fig.~\ref{fig:CF_spec} shows the resulting covering fraction evolution, where we identified the separation between the first and second eclipse. The lower panels show a selection of X-ray spectra and best-fitting models, all associated with the more significant first eclipse. 

\section{Cloud properties and size of the X-ray-emitting region}
\label{sec:properties}

The results from the X-ray time-resolved spectral analysis in Sect.~\ref{sec:time_resolved} can be used to derive estimates of the absorber and X-ray-emitting regions properties. We made a series of standard simplifying assumptions following earlier work, for instance by \citet{Lamer2003,Risaliti2005,Risaliti2007,Sanfrutos2013,Markowitz2014,Lian2025}, and we derive estimates of the eclipsing cloud size, density, transverse velocity, and location as well as of the X-ray-emitting region size. We focus here on the first eclipse only due to its much better sampling.

The proper characterisation of an X-ray eclipse requires the knowledge of the covering fraction and column density evolution throughout the event (and, possibly, of the ionisation state as well). However, the quality of the XRT data is insufficient to constrain simultaneously all quantities in any given spectrum and the whole eclipse was described by one single column density and ionisation parameter. We then make a series of simplifying assumptions that enable us to derive estimates of the cloud and X-ray source properties, which have, however, to be considered as estimates rather than precise measurements. We followed the procedure described by \citet{Sanfrutos2013} which is similar to that of many previous works \citep[e.g.][]{Lamer2003,Risaliti2007}. 

In particular, we considered a central eclipse by a spherical cloud with diameter $d_{\rm c}$, uniform density $n_{\rm c}$, and ionisation $\xi_{\rm c}$ moving along a Keplerian circular orbit around the SMBH with (transverse) velocity $v_{\rm c}$. X-rays are emitted uniformly from a circular region with diameter $d_{\rm X}$ coplanar with the accretion disc. Under these simplifying assumptions, the fact that the maximum covering fraction is $C_{\rm f, max} < 1$ (Fig.~\ref{fig:CF_spec}) implies $d_{c} < d_X$. The measured $N_{\rm H}$ (see Table~\ref{tab:table4}) is mostly constrained by the more absorbed spectra when the line-of-sight passes through the whole cloud, so 
\begin{align} 
N_{\rm H} \simeq d_{\rm c} n_{\rm c} \label{eq:NH}~.
\end{align}

\begin{figure}[t]
\centering 
\includegraphics[width=0.95\columnwidth]{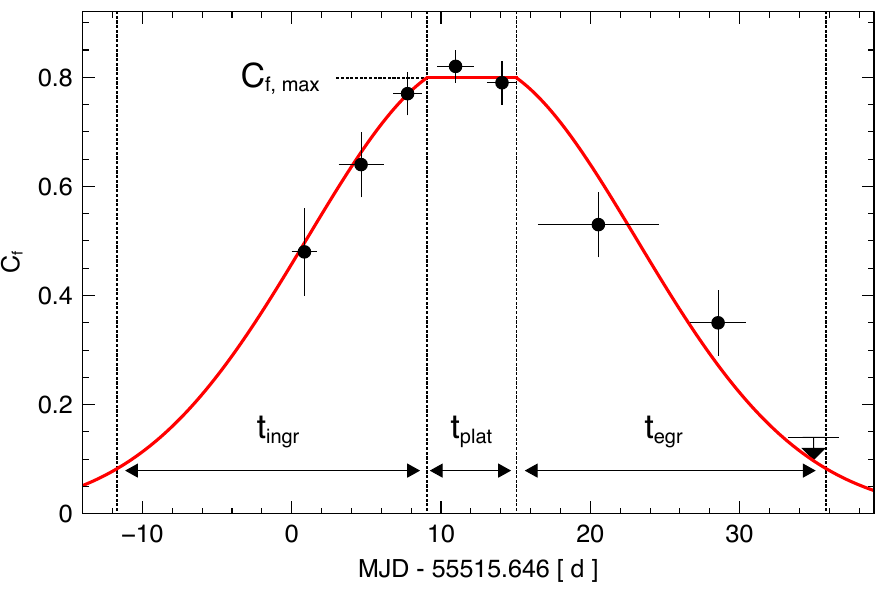}
\caption{Evolution of the covering fraction during the first eclipse. We also show a truncated (flat-top) Gaussian model example, together with the definition of the eclipse timescales (see text for details).}
\label{fig:CF_fits}
\end{figure}

The $C_{\rm f}$ evolution can be described by its maximum value $C_{\rm f, max}$ and by the following timescales: the ingress ($t_{\rm ingr}$) and egress ($t_{\rm egr}$) timescales are the time intervals during which $C_{\rm f}$ rises from zero to $C_{\rm f, max}$ and decays from $C_{\rm f, max}$ to zero respectively; the plateau timescale ($t_{\rm plat}$) is the time interval during which $C_{\rm f} = C_{\rm f, max}$. The sum of the three timescales is the total eclipse duration $t_{\rm tot}$. For a uniform-density cloud and a uniformly emitting X-ray source, the evolution is expected to be symmetric, so that $t_{\rm egr} = t_{\rm ingr}$ and $t_{\rm tot} = 2~t_{\rm ingr} + t_{\rm plat}$. These timescales are represented in Fig.~\ref{fig:CF_fits} where we show the $C_{\rm f}$ evolution for the first eclipse together with a simple truncated (flat-top) Gaussian model example (discussed below).

For $d_{\rm c}\leq d_{\rm X}$, the cloud diameter $d_{\rm c}$ and the relation between $d_{\rm c}$ and the diameter of the X-ray-emitting region $d_{\rm X}$ follows from the fact that the cloud transverse velocity $v_{\rm c}$ can be expressed in terms of the ingress (or egress) timescale $t_{\rm ingr}$  and the total eclipse duration $t_{\rm tot}$ as
\begin{align} 
v_{\rm c} & = \frac{d_{\rm c}}{t_{\rm ingr}} = \frac{d_{\rm X}+d_{\rm c}}{t_{\rm tot}} \label{eq:v}~.
\end{align}

On the other hand, the ionisation parameter associated with the ionised absorber model is defined as
\begin{align} 
\xi & = \frac{L_{\rm ion}}{n_{\rm c}R_{\rm c}^2} \label{eq:xi}~,
\end{align}
where $L_{\rm ion} =(2.6\pm 0.3) \times 10^{43}\ \mathrm{erg\ s^{-1}}$ is the averaged luminosity over the eclipse between $13.6$~eV and $13.6$~keV that was estimated from the joint spectral analysis of the XRT and UVOT UV filters data using the \texttt{agnsed} model \citep{KubotaDone2018} as described in Appendix~\ref{app:Lion}. 

Identifying $v_{\rm c}$ with the Keplerian velocity, the cloud density can be written as
\begin{align} 
n_{\rm c} \simeq \frac{N_{\rm H}}{d_c} = \frac{N_{\rm H}}{t_{\rm ingr}} \sqrt{\frac{R_{\rm c}}{GM_{\rm BH}}} \label{eq:n}~,
\end{align}
where $R_{\rm c}$ is the cloud distance from the central SMBH. Finally, by combining Eqs.~\ref{eq:xi} and \ref{eq:n} one has
\begin{align} 
R_{\rm c} & =  \left( GM_{\rm BH}\right)^{1/5}~\left( \frac{L_{\rm ion}~t_{\rm ingr}}{N_{\rm H}~\xi}\right)^{2/5}\label{eq:R}~,
\end{align}
which only depends on known quantities and on $t_{\rm ingr}$. 

To estimate $t_{\rm ingr}$ (as well as $C_{\rm f, max}$, $t_{\rm tot}$, and $t_{\rm plat}$) we initially considered a set of phenomenological models for the $C_{\rm f}$ evolution. We used piecewise functions with a plateau around the peak of $C_{\rm f}$, and considered exponential, Gaussian, and linear rise and decay. The plateau was modelled either with a constant or with a super-Gaussian\footnote{ $Ne^{-\frac{(t-t_0)^n}{2\sigma^n}}$ where $N$ is a normalisation, and $n$ the super-Gaussian order.} function of order $4$. The eclipse timescales turned out to be model-dependent and no unique solution was found, the most crucial parameter being $t_{\rm plat}$. However, the phenomenological models could be used to define the range of possible $t_{\rm plat} = [0,12]$~d.

We then adopted a flat-top symmetric Gaussian function where, in each fit, the plateau duration $t_{\rm plat}$ was fixed to a value in the range derived from the explored phenomenological models. However, we ignored $t_{\rm plat} = 0$~d as, for a central eclipse, this inevitably results in a maximum covering fraction $C_{\rm f, max} = 1$ which is inconsistent with the data (see Fig.~\ref{fig:CF_spec}). Since a fit with $t_{\rm plat} = 1$~d produces identical results to the case of $t_{\rm plat} = 0$~d, we also ignored the latter solution.

In summary, fits to the $C_{\rm f}$ evolution were performed with a flat-top Gaussian function with $t_{\rm plat} = [2,12]$~d in steps of $1$~d. We defined the total eclipse duration $t_{\rm tot}$ as the Gaussian full width at tenth of maximum, and we obtained $C_{\rm f, max}$, $t_{\rm tot}$, and $t_{\rm ingr} = 1/2\left(t_{\rm tot}-t_{\rm plat}\right)$ for each $t_{\rm plat}$. One realisation of such fits is shown in Fig.~\ref{fig:CF_fits} with $t_{\rm plat} = 6$~d. Table~\ref{tab:Gauss} reports the best-fitting values to the $C_{\rm f}$ evolution for $t_{\rm plat} = (2, 8, 12)$~d. All parameters for different $t_{\rm plat}$ are within the range of values defined by the extremes in $t_{\rm plat}$. 

The system of Eqs.~\ref{eq:NH}-\ref{eq:R} is closed, and a family of solutions (one for each $t_{\rm plat}$) for the cloud properties ($n_{\rm c}$, $d_{\rm c}$, $R_{\rm c}$, $v_{\rm c}$, and $d_{\rm c}$) can be derived. The X-ray-emitting region size $d_{\rm X}$ can then be obtained from $d_{\rm c}$, $v_{\rm c}$ and $t_{\rm tot}$ using Eq.~\ref{eq:v}. Table~\ref{tab:table2} reports results on all quantities of interest for $t_{\rm plat} = (2, 8, 12)~d$. Results for the two extremal values of $t_{\rm plat}$ represent upper and lower bounds for all parameters. Uncertainties were obtained by propagating all errors except for the black hole mass that was assumed to be $4.5\times 10^7~M_\odot$ \citep{Agis2014_ESO362}. For the cloud and X-ray-emitting size we adopt the natural symmetry and report $r=d/2$.

\begin{table}[t]
        \centering
        \caption{Best-fitting results for the $C_{\rm f}$ evolution using a flat-top Gaussian function. }
       \label{tab:Gauss}
        \begin{tabular}{lccc} 
\hline
\multicolumn{4}{c}{\T $C_{\rm f}$ evolution\B} \\
\hline
\T  $t_{\rm plat}$~[d] & $t_{\rm tot}$~[d] & $t_{\rm ingr}$~[d] & $C_{\rm f, max}$ \B \\
\hline          
\T  $2$ & $48.5\pm 3.6$ & $23.3\pm 1.8$ & $0.81\pm 0.02$ \B \\
\hline          
\T  $8$ & $46.5\pm 3.6$ & $13.3\pm 1.8$ & $0.79\pm 0.02$ \B \\
\hline          
\T  $12$ & $44.5\pm 3.6$ & $16.3\pm 1.8$ & $0.78\pm 0.02$ \B \\
\end{tabular}
\tablefoot{We only show results for $t_{\rm plat} =(2, 8, 12)$~d.}
\end{table}

\begin{table}[t]
        \centering
        \caption{Cloud and properties of the X-ray-emitting region for $t_{\rm plat} =(2, 8, 12)$~d. }
        \label{tab:table2}
        \begin{tabular}{lcccc}                                       
\T Parameter & $t_{\rm plat} = 2$~d &  $t_{\rm plat} = 8$~d & $t_{\rm plat} = 12$~d \B\\
\hline

\T $n_{\rm c}$~[$10^8$~cm$^{-3}$] &  $1.2\pm 0.3$ & $1.5\pm 0.4$ &$1.7\pm 0.5$ \B \\
\hline
\T $R_{\rm c}$~[$10^{17}$~cm] &  $1.9\pm 0.5$ & $1.7\pm 0.5$ & $1.6\pm 0.5$  \B \\
\hline
\T $v_{\rm c}$~[km~s$^{-1}$] &  $1790\pm 250$ & $1860\pm 265$ &$1925\pm 275$ \B \\
\hline
\T $r_{\rm c}^{\rm (a)}$~[$R_{\rm g}$] &  $27.1\pm 4.4$ & $23.3\pm 4.0$ & $20.3\pm 3.7$ \B \\
\hline
\T $r_{\rm X}^{\rm (a)}$~[$R_{\rm g}$] &  $29.4\pm 4.4$ & $33.0\pm 4.2$ & $35.3\pm 4.3$  \B \\
\hline
\end{tabular}
\tablefoot{
All quantities are bound by the values for the extremes in $t_{\rm plat}$.
We assumed an SMBH mass of $4.5\times 10^7~M_\odot$ so that $1~R_g \simeq 6.65\times 10^{12}$~cm. \\
$^{\rm (a)}$ Physically acceptable solutions for $r_{\rm c}$ and $r_{\rm X}$ can be constrained in a narrower range than shown here (see the text, Table~\ref{tab:tablef}, and Eq.~\ref{eq:rx}).
}
\end{table}

\begin{figure}[t]
\centering 
{\includegraphics[width=0.95\columnwidth]{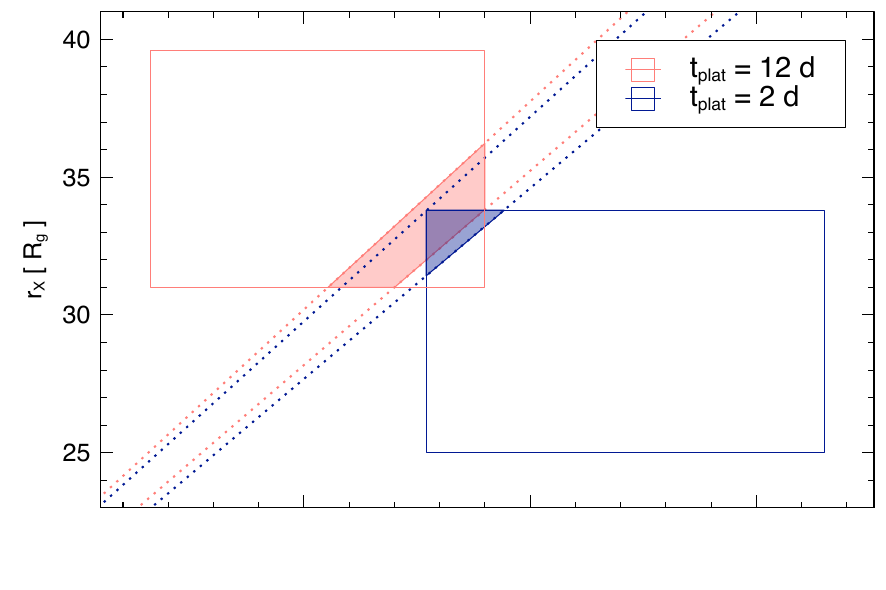}
\vspace{-0.2cm}
\hspace{0.02cm}
\includegraphics[width=0.95\columnwidth]{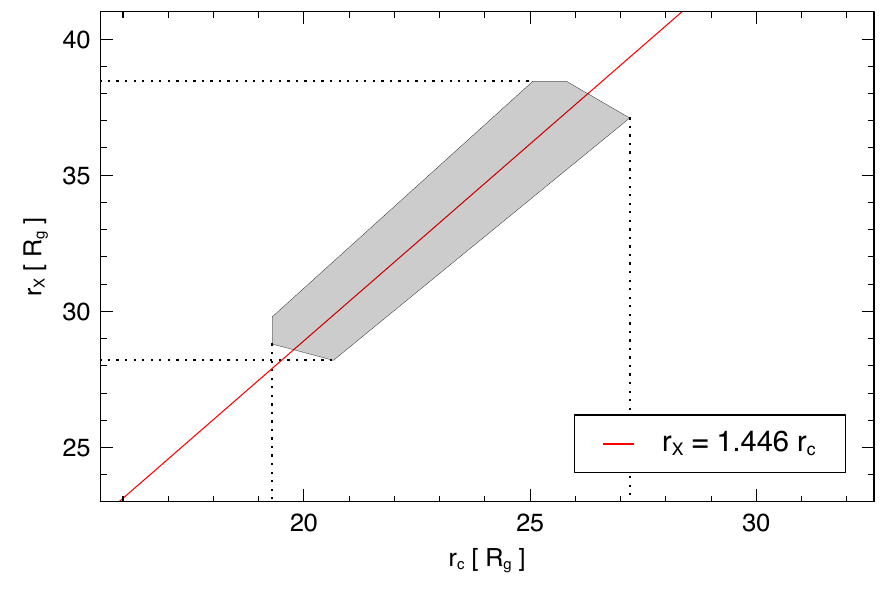}
}
\caption{{\it Top:} Allowed range of $r_{\rm c}$ and $r_{\rm X}$ that satisfy Eqs.~\ref{eq:NH}-\ref{eq:R} for the two extremal values of $t_{\rm plat} = 2$~d and $12$~d reported in Table~\ref{tab:table2} (two error boxes). Dotted lines denote Eqs.~\ref{eq:CFmax} for the two solutions, which differ because of slightly different $C_{\rm f, max}$. The shaded areas represent the restricted range of allowed $r_{\rm c}$ and $r_{\rm X}$ that satisfy Eqs.~\ref{eq:NH}-\ref{eq:R} and Eq.~\ref{eq:CFmax}. {\it Bottom:} Range of allowed solutions that satisfy Eqs.~\ref{eq:NH}-\ref{eq:R} and Eq.~\ref{eq:CFmax} for any $t_{\rm plat}$ in the range of $[2,12]$~d (shaded area). We also show the resulting averaged linear relation between $r_{\rm X}$ and $r_{\rm c}$ (solid red line). Dotted horizontal and vertical lines denote the overall range of $r_{\rm c}$ and $r_{\rm X}$.}
\label{fig:sizecomp}
\end{figure}

Table~\ref{tab:table2} shows that, considering uncertainties, the allowed cloud and X-ray-emitting region sizes span the broad ranges of $r_{\rm c} = [16.6,31.5]~R_{\rm g}$ and $r_{\rm X} = [25.0,39.6]~R_{\rm g}$. However, not all solutions within these ranges are allowed since some ($r_{\rm c}$, $r_{\rm X}$) pairs are inconsistent with the measured $C_{\rm f, max}$ (see Table~\ref{tab:Gauss}). Indeed, $C_{\rm f, max}$  was never used in our derivation, and since $r_{\rm c}$ and $r_{\rm X}$ must be consistent with it, the allowed range for $r_{\rm c}$ and $r_{\rm X}$ can be significantly narrowed. Considering a system inclination $i$, $C_{\rm f, max}$ corresponds to the ratio of the area of the cloud and the projected area of the X-ray source, that is, 
\begin{align} 
C_{\rm f, max} = \left(\frac{r_{\rm c}}{r_{\rm X}}\right)^2\frac{1}{\cos(i)}\label{eq:CFmax}~,
\end{align}
where $i$ is the line-of-sight inclination, estimated by \citet{Agis2014_ESO362} from relativistic disc reflection spectroscopy using high-quality \xmm\ data as $53^\circ \pm 5^\circ$. 

In the upper panel of Fig.~\ref{fig:sizecomp}, we show the error boxes of $r_{\rm c}$ and $r_{\rm X}$ for the two extremal values of $t_{\rm plat}$ in Table~\ref{tab:table2} . The dotted lines represent $r_{\rm X} \pm \delta r_{\rm X}$ as a function of $r_{\rm c}$ from Eq.~\ref{eq:CFmax}, where $\delta r_{\rm x}$ was derived by propagating the uncertainties on $i$ and $C_{\rm f, max}$. The range of ($r_{\rm c}$, $r_{\rm X}$) that simultaneously satisfy Eqs.~\ref{eq:NH}-\ref{eq:R} and Eq.~\ref{eq:CFmax} is significantly narrowed down with respect to that reported in Table~\ref{tab:table2} (which only comes from Eqs.~\ref{eq:NH}-\ref{eq:R}) and is shown as a shaded area for both solutions. 

By repeating this exercise for all values of $t_{\rm plat}$, the parameter space of the allowed ($r_{\rm c}$, $r_{\rm X}$) shrank significantly, as expected. In the lower panel, the shaded area shows the allowed solutions that simultaneously satisfy Eqs.~\ref{eq:NH}-\ref{eq:R} and Eq.~\ref{eq:CFmax} for any plateau duration within the range $t_{\rm plat} = [2,12]$~d. The dotted black lines show the overall allowed range of $r_{\rm X}$ and $r_{\rm c}$. As reference, we also provide an approximate linear relation that allowed us to derive $r_{\rm X}$ for any chosen value of $r_{\rm c}$ (respecting the overall ranges),

\begin{equation}
\label{eq:rx}
\begin{alignedat}{2}
r_{\rm X}\pm \Delta r_{\rm X}  &= r_{\rm c}~(\tilde{C}_{\rm f, max}\cos i)^{-1/2} \pm \Delta r_{\rm X} \\
& = 1.446~r_{\rm c} \pm 2.2~R_{\rm g}~,
\end{alignedat}
\end{equation}
where all quantities are in units of $R_{\rm g}$. $\tilde{C}_{\rm f, max}= 0.795$ is the median $C_{\rm f, max}$ for any $t_{\rm plat}$, $i=53^\circ$, and $\Delta r_{\rm X}$ is derived from the limits of the shaded area in the lower panel of Fig.~\ref{fig:sizecomp}. As per the cloud properties, they cannot be further constrained and all solutions in Table~\ref{tab:table2} remains valid. Table~\ref{tab:tablef} reports the final constraints on all properties where central values for $n_{\rm c}$, $R{\rm c}$, and $v_{\rm c}$ are the median of the allowed range for any $t_{\rm plat}$ with symmetric uncertainties.

\begin{table}[t]
        \centering
        \caption{Cloud and properties of the X-ray-emitting region for any $2~\mathrm{d} \leq t_{\rm plat} \leq 12$~d as constrained by Eqs.~\ref{eq:NH}-\ref{eq:R} and Eq.~\ref{eq:CFmax}.}
        \label{tab:tablef}
        \begin{tabular}{lccc} 
\hline
\T $n_{\rm c}$~[$10^8$~cm$^{-3}$] &&&  $1.55\pm 0.65$ \B \\
\hline
\T $R_{\rm c}$~[$10^{17}$~cm] &&&  $1.75\pm 0.65$ \B \\
\hline
\T $v_{\rm c}$~[km~s$^{-1}$] &&&  $1870\pm 330$ \B \\
\hline
\T $r_{\rm c}^{\rm (a)}$~[$R_{\rm g}$] &&&  $23.3\pm 4.0$ \B \\
\hline
\T $r_{\rm X}^{\rm (a)}$~[$R_{\rm g}$] &&&  $33.4\pm 5.2$ \B \\
\hline
\end{tabular}
\tablefoot{
We assumed an SMBH mass of $4.5\times 10^7~M_\odot$ so that $1~R_g \simeq 6.65\times 10^{12}$~cm.\\
$^{\rm (a)}$ Within the reported range of $r_{\rm c}$ and $r_{\rm X}$, the two quantities must satisfy Eq.~\ref{eq:rx}.
}
\end{table}

\section{One step further: Eclipse tomography of the inner accretion flow}
\label{sec:onestep}

We have so far assumed that the absorber affects simultaneously the whole X-ray spectrum whose main components are a soft X-ray excess, dominating below $\sim$~2~keV, and a hard power law component from a hot X-ray corona, dominating at higher energies. The analysis therefore implicitly assumes that the two spectral components originate from the same physical region. However, occultation events can potentially be used to gain insights on the actual geometry of the X-ray-emitting region(s). 

Some models postulate that the X-ray emission is radially stratified. The hot corona dominates the innermost radii out to $r_{\rm HC}$ and is replaced by the soft-excess-emitting region extending from $r_{\rm HC}$ out to $r_{\rm SE}$ \citep[e.g.][]{KubotaDone2018}. Depending on the vertical extent of the inner hot corona, the two X-ray-emitting regions (hot corona and soft excess) might appear as partially superimposed due to projection effects, especially at relatively high inclination, as is the case in ESO~362-G18. However, the superposition can never be complete, so that the expectation from this type of models is that $C_{\rm f, max}^{\rm(SE)} < C_{\rm f, max}^{\rm(HC)}$ and that the overall eclipse lasts longer for the soft excess than the hot corona. 

On the other hand, the hot corona might be an extended structure elevated above the innermost accretion flow in a slab or wedge geometry \citep[e.g.][]{Poutanen1996,Gianolli2023}. In this case, if the soft excess were produced in the inner flow below the hot corona, the covering fraction evolution towards the two X-ray spectral components would depend on the actual degree of superposition and on the hot corona patchiness and might even be identical. 

To study whether the \swift\ XRT data can provide some insights on the geometry of the two main X-ray components, we considered a further spectral model in which the covering fraction towards the soft excess (the \texttt{comptt} spectral model) and the hot corona (\texttt{nthcomp}) are not forced to always be identical. We then re-fitted the 12 X-ray spectra allowing the covering fraction towards the two spectral components to be different. Details on the spectral analysis are reported in Appendix~\ref{app:time-resolved-spec_II}. 

We found that the two covering fractions were consistent with being the same within uncertainties ($90$\% credible intervals) in eight out of twelve spectra. However, this was not the case in four of the eight spectra that define the first eclipse. Our final best-fitting model (hereafter Model~2) was therefore one in which four of the eight X-ray spectra during the first eclipse are characterised by different covering fractions for the soft excess ($C_{\rm f}^{\rm (SE)}$) and the hot corona ($C_{\rm f}^{\rm (HC)}$), and best-fitting results are reported in Table~\ref{tab:table5}. We obtained $C=5793$ for $6519$ degrees of freedom, to be compared with  $C=5819$ for $6523$ degrees of freedom that was obtained using Model~1 (where $C_{\rm f}^{\rm (SE)} = C_{\rm f}^{\rm (HC)}$ always). The AIC is lower for Model~2 with $\Delta \rm{AIC} = 17.9$, significantly higher than the $\Delta \rm{AIC} = 10$ that can be associated with a strong preference for a model over the other. Although the AIC method has some limitations \citep[see e.g.][]{Buchner2014}, we consider the improvement significant enough to warrant further study, as detailed below.

\begin{figure}
\centering 
\includegraphics[width=0.95\columnwidth]{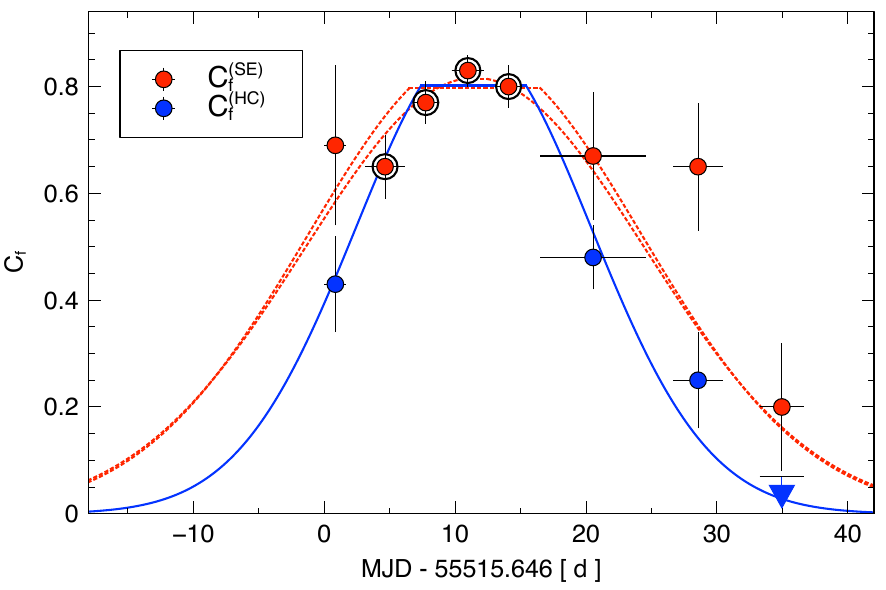}
\caption{Evolution of the covering fraction towards the soft excess (red) and hot corona (blue). Black circles denote  $C_{\rm f}^{\rm (SE)} = C_{\rm f}^{\rm (HC)}$. We also show flat-top Gaussian model-examples fitted to the $C_{\rm f}$ evolution of both components; see the main text for details.}
\label{fig:2CF}
\end{figure}

The resulting $C_{\rm f}$ evolution is shown in Fig.~\ref{fig:2CF} for the first eclipse only\footnote{For the second eclipse,  $C_{\rm f}^{\rm (SE)} = C_{\rm f}^{\rm (HC)}$ always (see Table~\ref{tab:table5}) and the common $C_{\rm f}$ evolution is equivalent to that shown in the upper panel of Fig.~\ref{fig:CF_spec}}. The soft excess covering fraction $C_{\rm f}^{\rm (SE)}$ is shown in red, and the hot corona one $C_{\rm f}^{\rm (HC)}$ in blue. Black circles denote data points (spectra) for which $C_{\rm f}^{\rm (SE)} = C_{\rm f}^{\rm (HC)}$. 

By applying a series of phenomenological models to both evolutions (as done in Sect.~\ref{sec:properties}), we could constrain the range of possible plateau duration for $C_{\rm f}^{\rm (HC)}$ and $C_{\rm f}^{\rm (HC)}$ as $t_{\rm plat}^{\rm (HC)}=[2,12]$~d and $t_{\rm plat}^{\rm (SE)}=[2,20]$~d. In Fig.~\ref{fig:2CF}, we show some examples of flat-top Gaussian models fitted to the $C_{\rm f}$ evolution. For the hot corona, the model corresponds to $t_{\rm plat}^{\rm (HC)} = 8$~d (blue solid line), while for the soft excess we show models corresponding to $t_{\rm plat}^{\rm (SE)} = 2$~d and $t_{\rm plat}^{\rm (SE)} = 10$~d (red dotted lines) to highlight that the soft-excess eclipse total duration is basically independent of the actual $t_{\rm plat}^{\rm (SE)}$. 

Although the models shown in Fig.~\ref{fig:2CF} have to be considered as simple representative examples, some indications can be be nevertheless obtained.\\
(I): For any $t_{\rm plat}$, the maximum covering fraction towards the two spectral components is reached approximately at the same time with $C_{\rm f, max}^{\rm (SE)} \simeq C_{\rm f, max}^{\rm (HC)}$ (see also the upper panel of Fig~\ref{fig:ProbDensity.pdf}); \\
(II): For any $t_{\rm plat}$, the eclipse lasts longer for the soft excess than the hot corona, that is $t_{\rm tot}^{\rm (SE)} > t_{\rm tot}^{\rm (HC)}$; \\
(III): Lastly, there is a clear outlier in the $C_{\rm f}^{\rm (SE)}$ evolution  for which $C_{\rm f}^{\rm (SE)}$ is significantly higher than predicted by a symmetric model (second to last data point in Fig.~\ref{fig:2CF}).

Points I and II differ under the simplifying assumption of uniformly emitting regions. A longer eclipse for the soft excess (point II) suggests a more extended emitting region than that of the hot corona. For a given cloud, in the simplest case of uniformly emitting regions, this would generally imply $C^{(\mathrm{SE})}_{f,\max} < C^{(\mathrm{HC})}_{f,\max}$, which is not what we observe. However, a centrally peaked emissivity profile in either component could reduce this difference, since the occultation of the brightest inner regions may dominate the effective covering fraction. Point III will be discussed later, in Sect.~\ref{sec:2regions}.

\subsection{A structured cloud: Dense core and tenuous atmosphere}

The apparent inconsistency between points I and II cannot be explained by geometrical effects alone. It might instead reflect the inadequacy of some simplifying assumption that we have made for the cloud structure. We present a plausible solution based on a structured cloud with denser core and larger, more tenuous atmosphere. 

\begin{figure}
\centering 
\includegraphics[width=0.95\columnwidth]{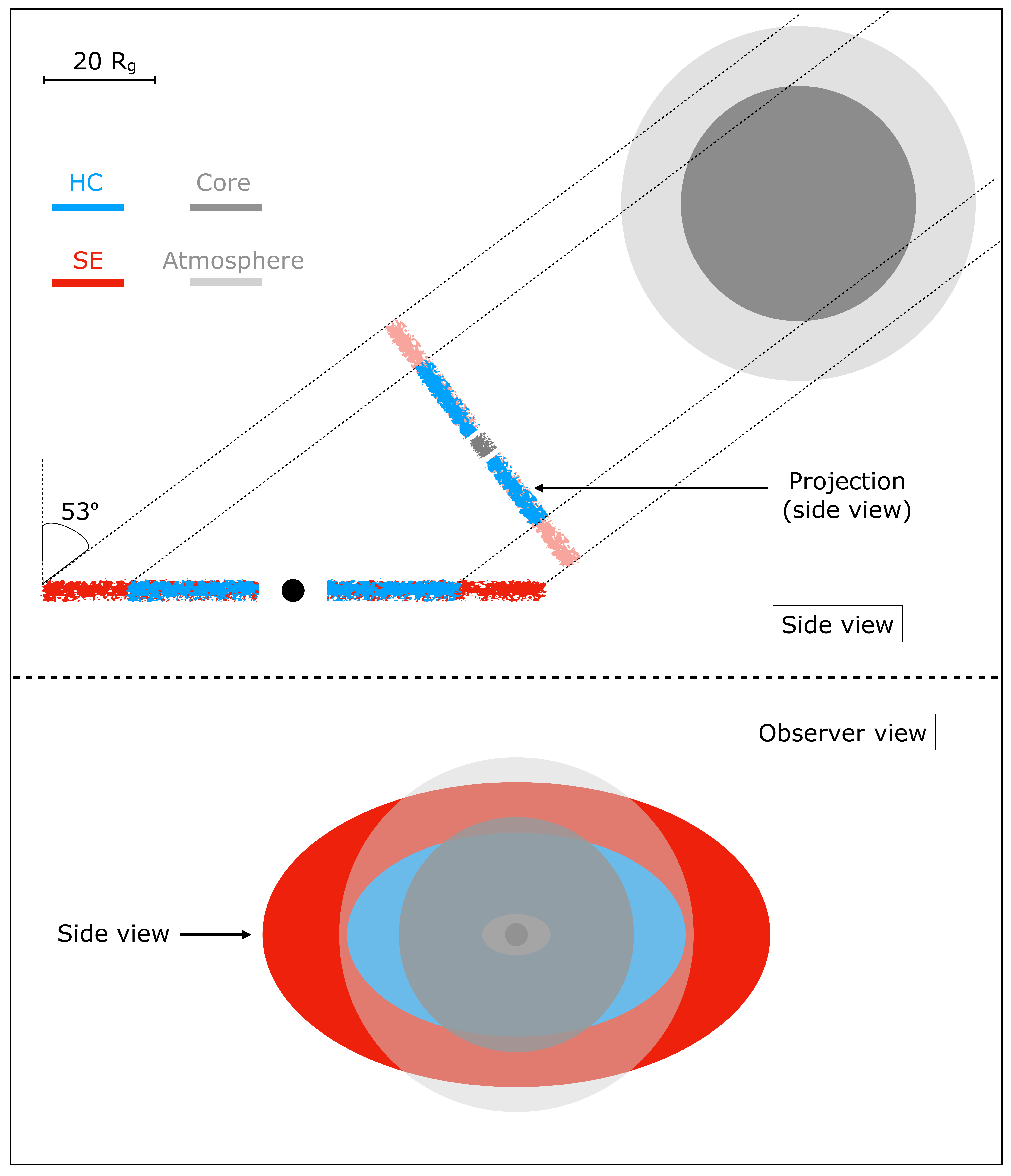}
\caption{Sketch of the eclipse geometry by a structured cloud. \textit{Top}: Side view of the system. \textit{Bottom}: Observer view for a line of sight inclined $53^\circ$ with respect to the accretion flow angular momentum direction. All structures are to scale using results in Table~\ref{tab:tabfinal}, although they only represent one of the possible solutions within the allowed ranges of sizes. The cloud also lies at much larger distance from the centre than shown. }
\label{fig:structured}
\end{figure}

The soft excess is significantly more sensitive to column density and ionisation variation than the hot corona which dominates above $\sim 2$~keV. Hence, even if the two emitting regions were co-spatial, the occultation of the soft excess by the cloud atmosphere would start earlier and last longer than that of the hot corona if the latter were unaffected by the tenuous atmosphere. A sketch of the geometry, for spherical cloud core and atmosphere, is shown in Fig.~\ref{fig:structured} from a side (upper panel) and from the observer point of view (lower panel). 

Assuming that the hot corona is only affected by the cloud core and considering that $v_{\rm atm} = v_{\rm core}$, Eq.~\ref{eq:v} leads to
\begin{equation}
\label{eq:sizerel}
\begin{alignedat}{2}
&r_{\rm SE} + r_{\rm atm} = (r_{\rm HC} + r_{\rm core})~\frac{t_{\rm tot}^{\rm (SE)}}{t_{\rm tot}^{\rm (HC)}} \\
&r_{\rm atm} = r_{\rm core}~\frac{t_{\rm ingr}^{\rm (SE)}}{t_{\rm ingr}^{\rm{(HC)}}}~. 
\end{alignedat}
\end{equation}
The core and hot corona properties can be derived as in Sect.~\ref{sec:properties} since, by assumption, the hot corona is not significantly affected by the cloud atmosphere. On the other hand, from the range of possible $t_{\rm plat}^{\rm (HC)} = [2,12]$~d and $t_{\rm plat}^{\rm (SE)}=[2,20]$~d, one has $t_{\rm tot}^{\rm (SE)}/t_{\rm tot}^{\rm (HC)} = 1.5\pm 0.2$, and $t_{\rm ingr}^{\rm (SE)}/t_{\rm ingr}^{\rm (HC)} = 1.5\pm 0.5$. Considering error propagation and imposing Eq.~\ref{eq:sizerel}, this translates into estimates for the size of the structured cloud and of the two X-ray-emitting regions that are reported in Table~\ref{tab:tabfinal}. 

\begin{table}[t]
        \centering
        \caption{Structured cloud and properties of the X-ray-emitting regions.}
        \label{tab:tabfinal}
        \begin{tabular}{lccc} 
\hline
\T $r_{\rm core}$~[$R_{\rm g}$] &&&  $19.5\pm 3.4$ \B \\
\hline
\T $r_{\rm atm}$~[$R_{\rm g}$] &&&  $29.3^{+11.0}_{-6.4}$ \B \\
\hline
\T $r_{\rm HC}$~[$R_{\rm g}$] &&&  $29.8\pm 3.7$ \B \\
\hline
\T $r_{\rm SE}$~[$R_{\rm g}$] &&&  $44.7^{+16.6}_{-11.2}$ \B \\
\hline
\end{tabular}
\end{table}

Our results suggest that the soft-excess-emitting region is about $50$\% more extended than the hot corona, although with relatively large uncertainties. In Table~\ref{tab:tabfinal}, the asymmetric uncertainties on $r_{\rm atm}$ and $r_{\rm SE}$ reflect the conditions $r_{\rm atm}>r_{\rm core}$ (by definition of a structured cloud) and $r_{\rm SE}>r_{\rm HC}$ (to prevent $C_{\rm f, max}^{\rm (SE)} > C_{\rm f, max}^{\rm (HC)}$ at maximum coverage by the cloud core, which is not observed).

It remains to be seen whether the cloud atmosphere can indeed absorb the soft X-rays without severely affecting the hard ones, as postulated. Assuming (arbitrarily) a factor of $5$ contrast between the core and cloud density, one has $n_{\rm atm} = 0.2~n_{\rm core} = (3.6\pm 1.4)\times 10^7$~cm$^{-2}$ and $\log\xi_{\rm atm} = \log(5~\xi_{\rm core}) = 1.34\pm 0.37$. The maximum atmosphere-only $N_{\rm H}$ can be evaluated at the core-atmosphere transition and is therefore $N_{\rm ~H, atm} \simeq 2 (r_{\rm atm}^2 - r_{\rm core}^2 )^{1/2}~n_{\rm atm} \simeq (1.0\pm 0.8)\times 10^{22}$~cm$^{-2}$. 

\begin{figure}
\centering 
\includegraphics[width=0.95\columnwidth]{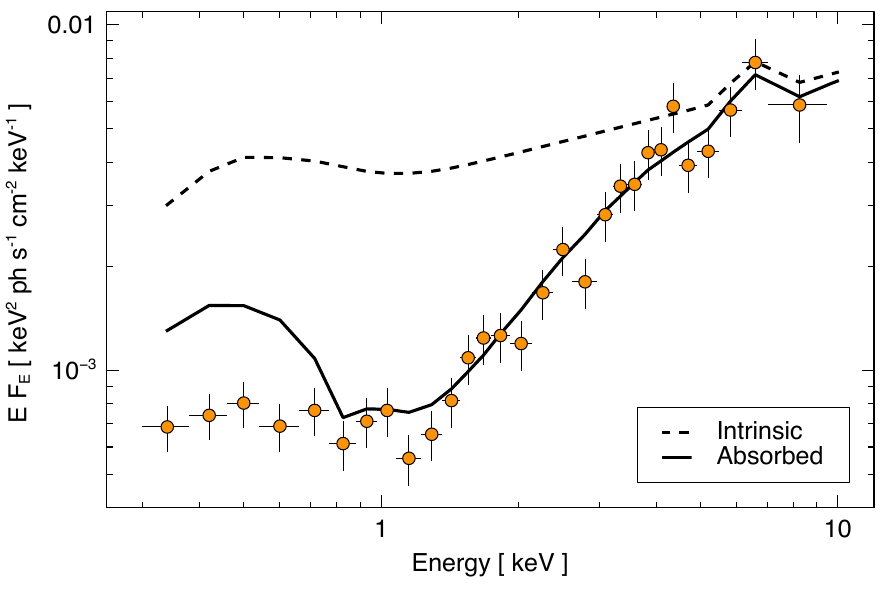}
\caption{Most absorbed spectrum (spectrum 4 in the upper panel of Fig.~\ref{fig:CF_spec}), shown together with its intrinsic unabsorbed spectral model (dashed) and a best-fitting model (solid) assuming a coverage of $80$\% ($20$\%) of the hot corona by the cloud core (atmosphere) and a coverage of  $0$\% ($100$\%) of the soft-excess-emitting region by the cloud core (atmosphere) with $N_{\rm ~H, atm} = 10^{22}$~cm$^{-2}$ and $\log\xi_{\rm atm} = 1.34$.}
\label{fig:spec4atm}
\end{figure}

To assess the maximum effect of such a cloud atmosphere on soft and hard X-rays we considered a simple power-law X-ray continuum model with $\Gamma = 2$ to which we applied one ionised absorber model with $C_{\rm f} = 1$ representing the cloud atmosphere. For the central values $N_{\rm ~H, atm} = 10^{22}$~cm$^{-2}$ and $\log\xi_{\rm atm} = 1.34$, the absorber induces a reduction of $\sim 5$\% of the hard X-ray flux above $2$~keV, while soft X-rays in the observed $0.3$-$2$~keV band are depressed eight times more efficiently by $\simeq 41$\%. The maximal absorption (obtained by adopting the higher and lower bounds in column density and ionisation respectively) reduces the hard X-ray flux by $\simeq 11$\% and the soft X-ray one by $\simeq 72$\%. Hence, the envisaged structured cloud appears to be able to account, at least qualitatively, for the observed behaviour and it represents a plausible explanation for the ($C_{\rm f}^{\rm (SE)}$, $C_{\rm f}^{\rm (HC)}$) evolution.

\subsection{Relative geometry of the two X-ray-emitting regions}

Finally, considering the geometry shown in Fig.~\ref{fig:structured}, it is interesting to assess whether the soft-excess-emitting region (red) is replaced by the hot corona (blue) in the innermost regions or rather co-exists with it. In the upper panels of Fig.~\ref{fig:geo}, we show a a schematic representation of these two possible geometries. In the case of radial stratification with no superposition between the two emitting regions - corresponding to case (b) in Fig.~\ref{fig:geo} -  at maximum coverage $\sim 80$\% of the hot corona is obscured by the core, and $\sim 20$\% by the atmosphere. On the other hand, the soft excess (which is only external to the hot corona), is basically only affected by the cloud atmosphere at maximum coverage. 

To assess whether this geometry is plausible, we considered the most absorbed spectrum (\#~4 in the upper panel of Fig.~\ref{fig:CF_spec}), and we re-fitted it with two ionised absorbers. The first absorber (cloud core) had column density and ionisation fixed to the best-fitting values derived earlier ($N_{\rm ~H, core} = 4.8\times 10^{22}$~cm$^{-2}$ and $\log\xi_{\rm core} = 0.64$; see Table~\ref{tab:table5}) and covered $80$\% of the hot corona without affecting the soft excess. The second (cloud atmosphere) had the column density and ionisation derived above  by assuming that $n_{\rm atm} = 0.2~n_{\rm core}$ ($N_{\rm ~H, atm} = 10^{22}$~cm$^{-2}$ and $\log\xi_{\rm atm} = 1.34$) and covered $20$\% of the hot corona and a fraction (free in the fit) of the soft-excess-emitting region. 

Figure~\ref{fig:spec4atm} shows the resulting best-fitting model, corresponding to full coverage of the soft excess by the cloud atmosphere. Even in that case, the cloud atmosphere alone is not sufficient to account for the depression of the soft X-ray flux at maximum coverage. Although the column density and ionisation of the atmosphere were set by the arbitrary choice of $n_{\rm atm} = 0.2~n_{\rm core}$, assuming a lower density contrast could increase $N_{\rm H, atm}$ and lower $\log\xi_{\rm atm}$ but at the expense of affecting significantly also the hard X-rays thus making the whole idea of a structured cloud ineffective. The only plausible solution is that the soft excess is also partially obscured by the cloud core. This means that the most likely solution for the geometry of the system is one in which the soft-excess-emitting region co-exists, at least partially, with the hot corona one in the innermost accretion flow rather than being replaced by it (panel (a) in Fig.~\ref{fig:geo}) and is therefore also partly obscured by the cloud core at maximum coverage.

\section{Discussion and conclusions}
\label{sec:discussion}

The time-resolved spectral analysis showed that the X-ray spectral variability of ESO~362-G18 during the \swift\ campaign was driven by two occultation events. The first event occurred within the initial $\simeq 40$~d and is more significant and better sampled than the second. Using data from the first eclipse, we were able to estimate the properties of the obscurer as well as the size of the X-ray-emitting region(s). The work presented here relied on a series of simplifying assumptions. Our results on (e.g.) X-ray-emitting region(s) sizes are therefore to be considered as estimates rather than precise measurements. Before we discuss our results, we recall the most important assumptions we made.

Based on the lack of significant spectral variability outside eclipse, we  assumed that the X-ray flux variability occurs at constant spectral shape. Subtle intrinsic spectral variability might affect our results, especially with respect to the composite X-ray-emitting region (soft excess and hot corona, as discussed in Sect.~\ref{sec:onestep}). We assumed a central eclipse for simplicity, and because the quality of the XRT data prevented us from simultaneously constraining column density and covering fraction, we assigned a single column density ($N_{\rm H} = dn$) to the eclipsing cloud. On the other hand, since the column density is mostly constrained by the most absorbed spectra, this simplification is not expected to have a major effect when the line of sight passes through the whole cloud (of size $d$). We considered uniformly emitting X-ray regions and ignored any radial emissivity profile as well as relativistic beaming and light bending. We also ignored the fact that no significant emission is expected within the innermost stable circular orbit.
The X-ray-emitting regions were also assumed to be circular and coplanar with the accretion disc: if the hot corona had a wedge-like or bi-conical geometry, its projection on the plane of the sky would be different than that of a coplanar region, depending on the wedge or cone opening angle, which can affect some of our results. We naturally assumed that the X-ray-emitting region(s) size(s) did not change during the observing campaign, which is not necessarily true, for example, if a relation exists between size and X-ray flux.  

\subsection{A single X-ray-emitting region}
\label{sec:1region}

By assuming a single uniformly emitting X-ray region, we derived in Sect.~\ref{sec:properties} that the first X-ray eclipse during the \swift\ campaign was consistent with being caused by a cloud with radius $r_{\rm c} \simeq 23.3~R_{\rm g}$ and density $n_{\rm c} \simeq 1.55\times 10^8$~cm$^{-2}$ (see Table~\ref{tab:tablef}). At Keplerian motion, the cloud transverse velocity is $v_{\rm c} \simeq 1870 $~km~s$^{-1}$, leading to a distance $R_{\rm c} = (1.75\pm 0.65)\times 10^{17}~\mathrm{cm}=(0.057\pm 0.021)$~pc from an SMBH with mass $M_{\rm BH} = 4.5\times 10^7~M_\odot$. The dust-sublimation radius in ESO~362-G18 is $R_{\rm dust} \simeq 0.14~{\rm pc} \simeq 4.3\times 10^{17}\mathrm{cm}$, where we adopted the definition by \citet{Rdust} for a sublimation temperature of $1500$~K and used the estimated $L_{\rm bol} \simeq 1.2\times 10^{44}$~erg~s$^{-1}$ of ESO~362-G18 (see Appendix~\ref{app:Lion}). The cloud distance $R_{\rm c} \simeq (0.41\pm 0.15)~R_{\rm dust}$ is therefore consistent with the innermost dust sublimation zone, which is typically defined as ${\rm DSZ} =0.4$-$1~R_{\rm dust}$. In other words, the cloud is located close to the boundary between the innermost dust-free zone (e.g. the broad-line region) and the outer dusty torus of Unification models \citep{Antonucci1993_unification}, confirming the clumpy nature of the circumnuclear medium.

The size of the X-ray-emitting region is constrained in the overall range of $r_{\rm X} = 33.4\pm 5.2~R_{\rm g}$. However, the estimates on ($r_{\rm c}$, $r_{\rm X}$) can be further refined by using Eq.~\ref{eq:rx}, accounting for the observed $C_{\rm f, max}$. Our results thus indicate a relatively compact but extended X-ray-emitting region. Another X-ray eclipse-like event in ESO~362-G18 was reported by \citet{Agis2014_ESO362}, although with much lower sampling. The authors were nevertheless able to estimate $r_{\rm X} \leq 48~R_{\rm g}$, which is fully consistent with the results reported here.  

\begin{figure*}[t]
\centering 
\includegraphics[width=1.9\columnwidth]{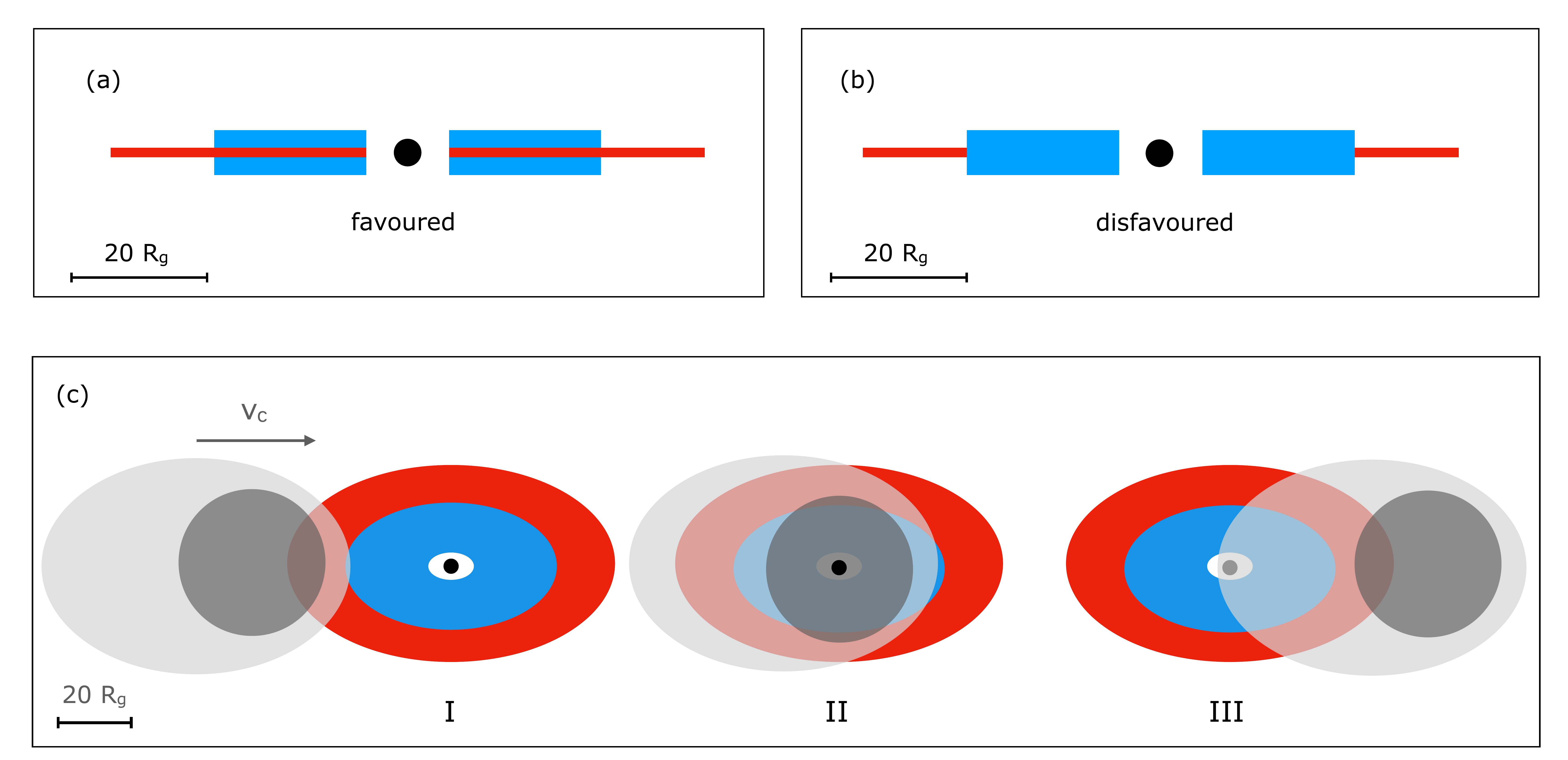}
\caption{{\it Panels a and b:} Schematic representation of two possible geometries for the hot corona (blue) and soft-excess-emitting (red) regions. Our analysis favours a geometry in which the soft excess and hot corona co-exist in the inner region rather than one of pure radial stratification (which cannot, however, be firmly ruled out).  {\it Panel c:} Schematic illustration of the occultation event. The drawing is approximately to scale, although the various structures sizes only represent one possible choice among the allowed ranges (see Table~\ref{tab:tabfinal}). The soft excess and hot corona emitting regions co-exist at the innermost radii and they  appear as ellipses due to projection effects ($i=53^\circ$). The cloud core and atmosphere are represented as dark and light grey areas, respectively. The cloud motion proceeds left to right, as indicated by its velocity vector, and three eclipse phases are identified (see the main text for details).}
\label{fig:geo}
\end{figure*} 

We also note that the AGN is characterised by transitions (three between 2003 and 2016) between optical spectral types 1.5 and 1.9, which classify ESO~362-G18 as a recurrent changing-look AGN. The relatively high line-of-sight inclination $i=53^\circ \pm 5^\circ$ \citet{Agis2014_ESO362} likely grazes the clumpy torus and thus enhances the probability of transient X-ray eclipses. It might also account for the absorption-driven changing-look phenomenology. However, the cloud that caused the X-ray eclipse studied here is certainly too small to account for the optical transitions even if it were dusty because the optical broad-line region is far larger than the size of the X-ray source.

\subsection{A composite X-ray-emitting region: Soft excess and hot corona}
\label{sec:2regions}

We also conducted a more complex X-ray spectral analysis in which the cloud-covering fraction towards the soft excess and hot corona X-ray-emitting regions was not forced to always be the same, assessing whether the data quality is high enough to enable us to perform basic X-ray tomography (see Sect.~\ref{sec:onestep}). We were able to infer that while the maximum covering fraction is consistent with being reached at approximately the same time and with being the same for the two components, the soft excess eclipse lasts longer than the hot corona eclipse by about a factor of $\simeq 1.5$. These two results are inconsistent with each other under the assumption of a uniform-density cloud and uniformly emitting X-ray regions. We therefore considered the case of a structured cloud comprising a denser core and an outer more tenuous atmosphere, and we showed that the results can be interpreted within this framework, as the cloud atmosphere can absorb the soft X-ray band (soft excess) while leaving the hard X-ray band essentially unaffected (dominated by hot corona emission). 

The envisaged cloud properties resemble the composite cloud structure proposed by \citet{Maiolino2010comet} in the context of occultation events by BLR clouds. Using high-quality data, \citet{Maiolino2010comet} inferred a density contrast of about a factor of $3$-$7$ between the cloud head (or dense core) and a cometary-like tail, similar to the contrast we assumed here between the core and atmosphere (a factor of 5, chosen as a mere example). A density contrast of a few between the cloud and atmosphere leads to atmosphere properties that are broadly consistent with the ionised gas that is observationally identified with the generic term of a warm absorber \citep[e.g.][]{Blustin2005,Crenshaw2012,Laha2014_wax1,Laha2016_wax2}. As discussed by \citet{Maiolino2010comet}, the lower-density atmosphere might be the result of supersonic motion of the cloud core in the ambient intra-cloud medium, producing a bow shock upfront and a Mach cone (or tail) behind the cloud head. In ESO~362-G18, we do not observe occultations from the BLR, but rather from a clumpy medium located towards the innermost dust sublimation zone, and the intra-cloud medium might be identified with the warm absorber. 

As discussed in point III, at the end of Sect.~\ref{sec:onestep}, we pointed out that the $C_{\rm f}$ evolution for the soft-excess region appears to be slightly asymmetric, with one data point around day $30$~d that is not well described by a symmetric model (see Fig.~\ref{fig:2CF}). This asymmetry in egress, however, is only tentative because at least one data point between $-20$~d and the start of the campaign during ingress would be needed to secure this claim. If it is real, the asymmetry  might support an elongated atmosphere similar to the cometary tail proposed by \citet{Maiolino2010comet} for the BLR clouds, inducing a longer egress than ingress, potentially accounting for the outlier in the $C_f^{\mathrm{(SE)}}$ evolution discussed in Sect.~\ref{sec:onestep} (point III).

The size of the hot corona is only slightly smaller than in the case of one single X-ray-emitting region, while the soft-excess-emitting region is $50$\% larger than the hot corona region, although with large uncertainties. A radially stratified geometry in which the soft-excess-emitting region is replaced by the hot corona in the innermost accretion flow without superposition appears disfavoured because the cloud atmosphere alone does not depress the soft X-ray emission sufficiently at maximum coverage, suggesting that the soft-excess-emitting region is also obscured by the cloud core. A more plausible solution appears to be one in which the soft-excess-emitting region is slightly more extended than the hot corona, but the two co-exist in the innermost accretion flow, or at least in part of it. 

Fig.~\ref{fig:geo} schematically illustrates the geometry of the X-ray-emitting regions and the proposed eclipse configuration. In the upper panels (a and b), we show two different geometries for the soft excess (red) and hot corona (blue), and as mentioned, our analysis suggests that a solution in which the two components co-exist in the inner region, shown in panel (a), is favoured. 

In the lower panel of Fig.~\ref{fig:geo} (c), we show the proposed eclipse time-evolution, where we assumed a cometary-like shape for the cloud atmosphere, and we considered that the soft excess and hot corona co-exist in the innermost region.  We assumed coplanar and uniformly emitting regions, ignoring relativistic beaming and light bending, and only accounted for projection effects \citep[see e.g.][for a proper treatment of these effects]{Kammoun2018}. We show three phases of the eclipse. In phase I the denser core of the cloud does not yet obscure the hot corona, which is also effectively unaffected by the atmosphere because its column density is lower and the ionisation is higher. On the other hand, the soft-excess-emitting region is already partially obscured by the cloud core and atmosphere. Phase II corresponds to the time of maximum obscuration for both components with a similar covering fraction. In phase III, the hot corona is effectively unobscured, while the soft-excess-emitting region is still partially covered by the core and the atmosphere. The soft-excess egress (phase III) might last longer than the ingress (phase I) if the cloud geometry were cometary-like with an extended trailing tail. This cloud geometry potentially accounts for the outlier in the $C_{\rm f}^{\rm (SE)}$ evolution in Fig.~\ref{fig:2CF} (penultimate data point).

The proposed geometry implies that the soft excess and hot corona most likely co-exist in the innermost regions, with the soft excess extending farther out than the hot corona. If the latter is elevated above the accretion flow and that the soft excess originates from below, this means that part of the soft X-ray emission reaches the observer directly, and part of it passes through the hot corona. A fraction of the latter might go through almost unaffected (especially if the hot corona is somewhat patchy), and the remainder is instead Comptonised and up-scattered by the hot corona into higher-energy photons. If this is so, the soft-excess emission might contribute significantly to coronal cooling. This additional cooling would naturally imply that AGNs with stronger soft excess exhibit a softer hard X-ray continuum slope, as is generally observed \citep[e.g.][]{Noda2018,Chen2025,Jana2026J}. 

For the origin of the soft excess itself, our results are inconclusive because a warm corona and relativistic disc reflection can be consistent with the proposed geometry. However, the proposed co-existence of the two spectral components in the innermost regions suggests that reprocessing of the hard X-ray photons from the hot corona is inevitable, at least to some extent. Hence, we expect that the soft excess is at least partly due to reprocessing in the innermost $20$-$40~R_{\rm g}$ in ESO~362-G18, consistent with the detection of a broadened relativistic Fe line and of a soft X-ray lag, which was interpreted as due to reverberation by \citet{Agis2014_ESO362}.

\begin{acknowledgements}
We thank the anonymous referee for their helpful comments and suggestions, which have significantly improved the quality and clarity of this work. We acknowledge the use of public data from the \textit{Neil Gehrels Swift} Observatory data archive. This work made use of data supplied by the UK \swift\ Science Data Centre at the University of Leicester. We acknowledge the use of data and software provided by the High Energy Astrophysics Science Archive Research Center (HEASARC), which is a service of the Astrophysics Science Division at NASA/GSFC. Most figures have been produced using the \texttt{VEUSZ} plotting package by Jeremy Sanders and contributors. This work was mostly realised during an internship of LRN at the Centro de Astrobiolog\'ia building upon and extending work included in BAG's PhD thesis. BAG is funded by the European Union ERC-2022-STG - BOOTES - 101076343. Views and opinions expressed are however those of the author only and do not necessarily reflect those of the European Union or the European Research Council Executive Agency. Neither the European Union nor the granting authority can be held responsible for them. GM acknowledges support from grants n. PID2020-115325GB-C31 and n. PID2023-147338NB-C21 funded by MICIU/AEI/10.13039/501100011033 and ERDF/EU.
\end{acknowledgements}

\bibliographystyle{aa}
\bibliography{biblio}


\begin{appendix}

\section{X-ray spectral analysis}
\label{app:Xrays_epoch1}

We present a detailed description of the X-ray spectral analysis. Additional information on the spectral grouping, error computation, and statistical model comparison is given in Sect.~\ref{sec:obs} and is not repeated here.

\subsection{Baseline spectral model}
\label{app:baseline_spec}

Previous high-quality X-ray observations of ESO~362-G18 have shown that the X-ray spectrum is that of a typical Seyfert~1 galaxy described, at zeroth-order, by a combination of a hard X-ray power law, a soft X-ray excess, and a reflection component mostly visible via Fe~K emission around 6.4~keV \citep{Agis2014_ESO362,Xu2021_ESO362,Zhong2022}. The soft excess can be interpreted as being due to either an optically thick ($\tau \gg 1$) warm ($kT_{\rm e} \sim 100$-$200$~eV) corona or a relativistically blurred reflection component off the inner accretion disc (or their combination). 

We adopted the simplest possible X-ray continuum model comprising a warm X-ray corona describing the soft excess and a hard thermal Comptonisation component from a hot optically thin corona. The reason for describing the soft X-ray excess through a warm corona rather than disc reflection is driven by model's simplicity, as reflected by the smaller number of free parameters which also results in fewer degeneracies between the spectral parameters. Moreover, many of the parameters associated with relativistic reflection cannot be efficiently constrained from the \swift\ XRT spectra alone, so that our choice appears to be better suited to the overall XRT dataset.

The soft excess was modelled using the \texttt{comptt} model in \texttt{xspec} \citep{Titarchuk1994_comptt} assuming a disc-like seed photons distribution with temperature of $2$~eV, consistent with standard accretion disc theory for the SMBH mass ($\sim 4.5\times 10^7~M_\odot$) and typical Eddington ratio ($\lambda \sim 0.02$) of ESO~362-G18 \citep[see also][]{Petrucci2018}. The hard Comptonisation component was described with the \texttt{nthcomp} model \citep{Zdziarski1996_nthcomp,Zycki1999_nthcomp} where we assumed the same seed photon distribution as for the soft excess. Due to the lack of hard X-ray data above $10$~keV, the electron temperature in \texttt{nthcomp} could not be constrained, and we fixed it to a standard value of $100$~keV. We also included a Gaussian emission line with fixed rest-frame energy at $\sim 6.4$~keV and width $\sigma_{\rm Fe}$ associated with Fe~K$\alpha$ emission. The overall model was absorbed by the Galactic column density fixed at $1.35\times 10^{20}$~cm$^{-2}$ \citep{HI4PI2016} described by the \texttt{tbabs} model using cross sections and abundances from \citet{Wilms2000_abund}. 

\subsection{Low and high H/S spectra}
\label{app:low_high_HSspec}

We first considered the low and high H/S spectra shown in Fig.~\ref{fig:HSspec} and corresponding to the time intervals shown in blue (low H/S) and red (high H/S) in Fig.~\ref{fig:epoch1_rebinned}, fitting them simultaneously with the model described above. The model provided a good description of the two spectra with $C=1234$ for $1320$ degrees of freedom. However, to describe the observed spectral variability, the photon index of the hard Comptonisation component was found to vary from $\Gamma \simeq 1.74$ in the low H/S state to an unphysical $\Gamma \lesssim 1.05$ in the high H/S state. We therefore rejected this solution on physical rather than statistical grounds.

As discussed in Sect.~\ref{sec:specvar} (and Fig.~\ref{fig:HS_1}), the intrinsic X-ray variability appears to occur at fixed spectral shape (H/S$\simeq 0.22$) indicating that the observed H/S variability is extrinsic and most likely due to intervening absorption, as also clear from the shape of the two X-ray spectra shown in Fig.~\ref{fig:HSspec}. We then added a layer of neutral absorbing gas local to the source, described by the \texttt{ztbpcf} model in \texttt{xspec}, with common column density and variable covering fraction $C_{\rm f}$ between the two spectra, and we forced all continuum parameters to be the same except for an overall normalisation. In other words, we attempted to describe the low and high H/S spectra with the simplest possible model in which the only variable parameters are an overall normalisation (thus preserving the intrinsically constant H/S$\sim 0.22$ in Fig.~\ref{fig:HS_1}) and the covering fraction of the intervening absorber, all other parameters being the same for both spectra. We obtained a good description of the two datasets that resulted in $C=1220$ for $1319$ degrees of freedom. No statistically significant improvement was obtained by letting any other parameter free to vary independently in the two spectra.

The (common) hard photon index was constrained to be $\Gamma = 1.74 \pm 0.07$, in line with the typical spectral shape of AGNs above $\sim 2$~keV. The absorber column density was measured to be $\simeq 5\times 10^{22}$~cm$^{-2}$, while its covering fraction was consistent with zero in the low H/S spectrum and was $C_{\rm f} \simeq 0.76$ in the high H/S one. By replacing the neutral partial covering model with the ionised \texttt{zxipcf} one \citep{Reeves2008_zxipcf}, we obtained an improvement of $\Delta C= -14$ for one degree of freedom, for a final result of $C=1206$ for $1318$ degrees of freedom. The improvement corresponds to $\Delta {\rm AIC} = 11.96 > 10$, and we retained the ionised absorber into our best-fitting model. The most relevant best-fitting parameters are discussed in the main text in Sect.~\ref{sec:low_high_HS}, and the best-fitting models and resulting residuals are shown in Fig.~\ref{fig:HSspec}.

\subsection{Time-resolved spectroscopy~I}
\label{app:time-resolved-spec}

Having established that the spectral difference between the low and high H/S spectra can be entirely attributed to a difference in absorption, we consider time-resolved spectral analysis for the 12 X-ray spectra corresponding to the data points shown in Fig.~\ref{fig:epoch1_rebinned}, each representing the stack of three consecutive \swift/XRT observations. The shape of the H/S light curve in Fig.~\ref{fig:epoch1_rebinned}, combined with results from the spectral analysis of the low and high H/S spectra, strongly suggests that the variable H/S during the initial $\sim 40$~d can be attributed to a single occultation (or eclipse) that progressively covers and uncovers the X-ray-emitting region, with a possible further eclipse around $50$-$60$~d. 

\begin{table}[t!]
        \centering
        \caption{Model~1: Best-fitting model parameters assuming a single X-ray-emitting region.}
        \label{tab:table4}
        \begin{tabular}{lccc}                                      
\hline
\multicolumn{4}{c}{\T Continuum and Fe line \B } \\
\hline 
\T Spec.~\# & $\Gamma$ & $kT_{\rm e}^{\rm (hot)}$ & $kT_{\rm e}^{\rm (warm)}$ \B \\
\hline
\T  1-12 & $1.71\pm 0.03$ & $100^f$ & $120\pm 10$ \B \\
\hline
\T  & $\tau^{\rm (warm)}$ & $\sigma_{\rm Fe}$ & EW$_{\rm Fe}$  \B \\
\hline
\T  1-12 & $\geq 25$ & $480\pm 220$ & $220\pm 130$ \B \\
\hline

\multicolumn{4}{c}{\T First eclipse \B }  \\
\hline

\T Spec.~\# & \multicolumn{3}{c}{ $N_{\rm H}$ \hspace{1.6cm} $\log\xi$\B }  \\
\hline
\T 1-8 & \multicolumn{3}{c}{ \hspace{0.cm} $4.5\pm 1.0$ \hspace{0.65cm} $0.65\pm 0.35$ \B }  \\
\hline

\T & $C_{\rm f}$ & $F_{\rm{0.3-10}}^{\rm obs}$ & $F_{\rm{0.3-10}}^{\rm int}$ \B \\
\hline
\T 1 & $0.48\pm 0.08$ & $1.56\pm 0.09$ & $2.27\pm 0.24$ \B \\
\hline
\T 2 & $0.64\pm 0.06$ & $1.30\pm 0.09$& $2.21\pm 0.24$  \B \\
\hline
\T 3 & $0.77\pm 0.04$ & $1.25\pm 0.09$& $2.39\pm 0.25$  \B \\
\hline
\T 4 & $0.82\pm 0.03$ & $1.34\pm 0.09$& $2.73\pm 0.27$  \B \\
\hline
\T 5 & $0.79\pm 0.04$ & $1.02\pm 0.08$& $2.02\pm 0.23$  \B \\
\hline
\T 6 & $0.53\pm 0.06$ & $3.11\pm 0.15$& $4.65\pm 0.41$  \B \\
\hline
\T 7 & $0.35\pm 0.06$ & $3.45\pm 0.11$& $4.32\pm 0.31$ \B \\
\hline
\T 8 & $\leq 0.14$ & $4.46\pm 0.13$& $4.72\pm 0.13$  \B \\
\hline
\multicolumn{4}{c}{\T Second eclipse \B }  \\
\hline

\T Spec.~\# & \multicolumn{3}{c}{ $N_{\rm H}$ \hspace{1.6cm} $\log\xi$\B }  \\
\hline
\T 9-12 & \multicolumn{3}{c}{ \hspace{-0.2cm} $0.7\pm 0.5$ \hspace{1.1cm} $\leq 0.8$ \B }  \\
\hline

\T & $C_{\rm f}$ & $F_{\rm{0.3-10}}^{\rm obs}$ & $F_{\rm{0.3-10}}^{\rm int}$ \B \\
\hline
\T 9 & $\leq 0.08$ & $5.61\pm 0.13$& $5.80\pm 0.13$  \B \\
\hline
\T 10 & $0.15\pm 0.07$ & $4.29\pm 0.14$& $4.78\pm 0.39$  \B \\
\hline
\T 11 & $0.31\pm 0.06$ & $4.10\pm 0.13$& $4.89\pm 0.38$ \B \\
\hline
\T 12 & $\leq 0.08$ & $3.10\pm 0.07$& $3.22\pm 0.09$ \B \\
\hline
\multicolumn{4}{c}{\T Global fit \B }  \\
\hline
\T 1-12 & & & $C/\nu = 5819/6523$  \B  \\
\hline
\end{tabular}
\tablefoot{
Column densities are given in units of $10^{22}$~cm$^{-2}$. $kT_{\rm e}^{\rm (warm)}$, $\sigma_{\rm Fe}$, and EW$_{\rm Fe}$ are in units of eV while $kT_{\rm e}^{\rm (hot)}$ is fixed and in units of keV. The model is also absorbed by Galactic column density fixed at $1.35\times 10^{20}$~cm$^{-2}$ \citep{HI4PI2016}. Observed and intrinsic (unabsorbed) X-ray fluxes are in units of $10^{-11}$~erg~s$^{-1}$~cm$^{-2}$. The first column (Spec.~\#) indicates the spectrum or range of spectra from which the reported parameters have been constrained (see e.g. the upper panel of Fig.~\ref{fig:CF_spec} as reference). The continuum and Fe line parameters (upper part) are common for all spectra (Spec.~\#1-12).}
\end{table}

\begin{figure}[t]
\centering 
\includegraphics[width=0.95\columnwidth]{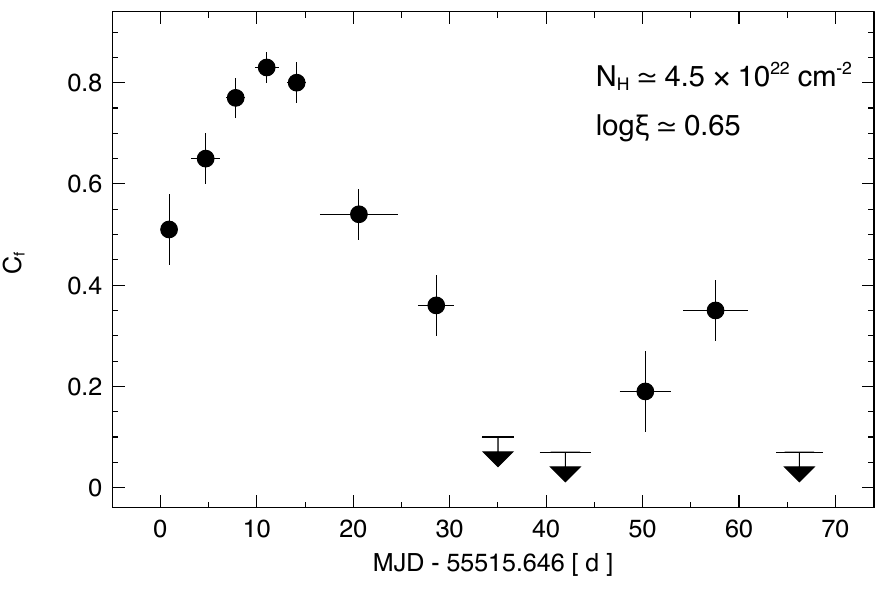}
\caption{Evolution of the covering fraction from the time-resolved spectroscopic analysis for common absorber properties ($N_{\rm H}$ and $\log\xi$) throughout the campaign.}
\label{fig:CF_commonNH}
\end{figure}

We therefore adopted the best-fitting model described above (see also Eq.~\ref{eq:specmodel}), fitting the 12 X-ray spectra simultaneously and allowing only an overall normalisation and the covering fraction $C_{\rm f}$ of the ionised absorber to vary independently. The 12 X-ray spectra are well described by the model with $C=5829$ for $6525$ degrees of freedom. The absorber's parameters are $N_{\rm H} \simeq 4.5 \times 10^{22}$cm$^{-2}$ and $\log\xi \simeq 0.65$; the soft excess has a plasma temperature $kT_{\rm e} = 120\pm 10 $~eV and poorly constrained optical depth $\tau \geq 25$; the hard Comptonisation slope is  $\Gamma = 1.72\pm 0.05$. The resulting covering fraction evolution is shown in Fig.~\ref{fig:CF_commonNH}. Two well-separated events are identified, suggesting the presence of two distinct X-ray eclipses in the data, the first encompassing the first eight data points, the second affecting the last four. Although the statistical quality of the fit was already excellent, it is more physically plausible that the two distinct events are characterised by different absorber properties (column density and ionisation). 

We therefore considered a further model in which the absorber properties  were allowed to vary independently between the two events. The statistical quality marginally improved by $\Delta C = -10$ ($C=5819$ for $6523$ degrees of freedom). Although the improvement cannot be considered as highly significant ($\Delta {\rm AIC} = 5.96$),  we retained this latter model as the best-fitting one since two events by two different absorbers are unlikely to be characterised by exactly the same properties\footnote{We have nevertheless checked that our results are unaffected by this choice, which has a negligible effect on the other relevant parameters}. The absorber has $N_{\rm H} = (4.5\pm 1.0) \times 10^{22}$cm$^{-2}$ and $\log\xi = 0.65 \pm 0.35$ during the first eclipse, and $N_{\rm H} = (0.7\pm 0.5) \times 10^{22}$cm$^{-2}$ with a poorly constrained $\log\xi \leq 0.8$ during the second. The best-fitting parameters are reported in Table~\ref{tab:table4} and the model is named Model~1.

\subsection{Time-resolved spectroscopy~II: Soft excess and hot corona}
\label{app:time-resolved-spec_II}

\begin{table}[t]
        \centering
        \caption{Model~2: Best-fitting model parameters when the soft excess and hot corona are allowed to be associated with different covering fraction ($C_{\rm f}^{\rm{(SE)}}$ and $C_{\rm f}^{\rm{(HC)}}$ respectively).}
        \label{tab:table5}
        \begin{tabular}{lcc}                                      
\hline 
\multicolumn{3}{c}{\T First eclipse \B }  \\
\hline
\T Spec.~\# & $N_{\rm H}$ & $\log\xi$ \B \\
\hline
\T 1-8 & $4.8\pm 1.0$ & $0.64\pm 0.37$ \B \\
\hline
\T & $C_{\rm f}^{\rm{(SE)}}$ & $C_{\rm f}^{\rm{(HC)}}$ \B \\
\hline
\T 1 & $0.69\pm 0.15$ & $0.43\pm 0.09$ \B \\
\hline
\T 2 & $0.65\pm 0.06$ & $=C_{\rm f}^{\rm{(SE)}}$ \B \\
\hline
\T 3 & $0.77\pm 0.04$ & $=C_{\rm f}^{\rm{(SE)}}$ \B \\
\hline
\T 4 & $0.83\pm 0.03$ & $=C_{\rm f}^{\rm{(SE)}}$ \B \\
\hline
\T 5 & $0.80\pm 0.04$ & $=C_{\rm f}^{\rm{(SE)}}$ \B \\
\hline
\T 6 & $0.67\pm 0.12$ & $0.48\pm 0.06$ \B \\
\hline
\T 7 & $0.65\pm 0.12$ & $0.25\pm 0.09$ \B \\
\hline
\T 8 & $0.20\pm 0.12$ & $\leq 0.07$ \B \\
\hline
\multicolumn{3}{c}{\T Second eclipse \B }  \\
\hline
\T  Spec.~\# & $N_{\rm H}$ & $\log\xi$ \B \\
\hline
\T 9-12 & $0.7\pm 0.5$ & $\leq 0.8$ \B \\
\hline
\T & $C_{\rm f}^{\rm{(SE)}}$ & $C_{\rm f}^{\rm{(HC)}}$ \B \\
\hline
\T 9 & $\leq 0.08$ & $=C_{\rm f}^{\rm{(SE)}}$ \B \\
\hline
\T 10 & $0.16\pm 0.07$ & $=C_{\rm f}^{\rm{(SE)}}$ \B \\
\hline
\T 11 & $0.32\pm 0.06$ & $=C_{\rm f}^{\rm{(SE)}}$ \B \\
\hline
\T 12 & $\leq 0.08$ & $=C_{\rm f}^{\rm{(SE)}}$ \B \\
\hline
\multicolumn{3}{c}{\T Global fit \B }  \\
\hline
\T 1-12 & & $C/\nu = 5793/6519$  \B  \\
\hline
\end{tabular}
\tablefoot{
Column densities are given in units of $10^{22}$~cm$^{-2}$. The X-ray continuum and Fe line parameters as well as X-ray fluxes are all consistent within errors with those reported in Table~\ref{tab:table4}. }
\end{table}

\begin{figure}[t]
\centering 
\includegraphics[width=0.95\columnwidth]{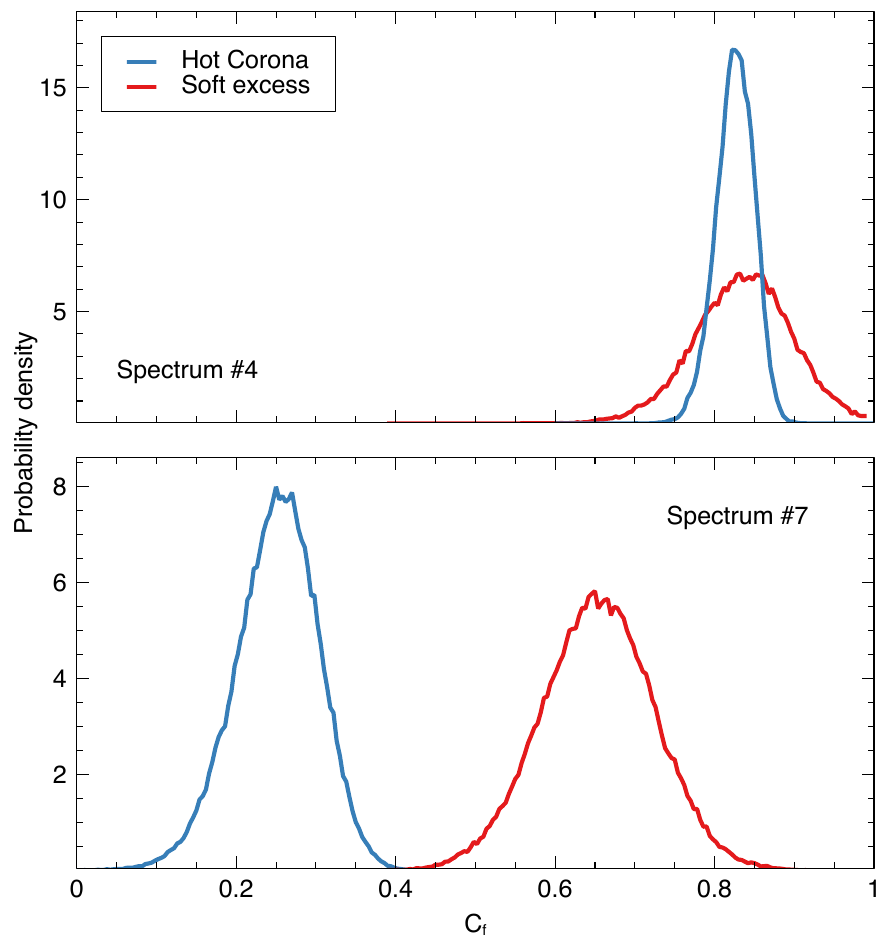}
\caption{Probability density for the covering fraction towards the soft excess and the hot corona in X-ray spectra \#~4 and \#~7.}
\label{fig:ProbDensity.pdf}
\end{figure}

To study whether the data can provide some insight on the actual geometry of the soft excess and hot corona emitting regions, we considered a further spectral model in which the covering fractions towards the soft excess and hot corona  X-ray-emitting regions ($C_{\rm f}^{\rm(SE)}$ and $C_{\rm f}^{\rm(HC)}$) were allowed to be different. After a few initial tests, we found that, in most cases (8 out of 12), the two covering fractions were consistent with being the same within uncertainties ($90$\% credible intervals). In our analysis, we then only retained as variable the remaining four $C_{\rm f}$ that are all associated with the first eclipse. As an example, in Fig.~\ref{fig:ProbDensity.pdf} we show the probability density associated with the two covering fractions for spectra \#~4 and \#~7. The two covering fractions in spectrum \#~4 (the most absorbed one) are fully consistent with each other and there is no justification for allowing them to be different. On the other hand, the opposite situation holds for spectrum \#~7. 

The data are well described  by this new model (Model~2) with $C=5793$ for $6519$ degrees of freedom with best-fitting parameters reported in Table~\ref{tab:table5}. The AIC for Model~1 and 2 implies a difference $\Delta {\rm AIC} = 17.9$ suggesting a strong preference for Model~2 over Model~1. Notwithstanding some limitations of the AIC method \citep[e.g.][]{Buchner2014}, the measured $\Delta {\rm AIC}$ is high enough that Model~2, in which $C_{\rm f}^{\rm(SE)}$ and $C_{\rm f}^{\rm(HC)}$ are not forced to always be identical, is worth considering. The implications of this analysis are discussed in Sect.~\ref{sec:onestep}.

\subsection{Ionising luminosity, $L_{\rm ion}$}
\label{app:Lion}

The ionisation parameter in Eq.~\ref{eq:xi} depends on the ionising luminosity $L_{\rm ion}$, that is the intrinsic AGN luminosity integrated from $13.6$~eV to $13.6$~keV. $L_{\rm ion}$ can be estimated from the bolometric luminosity $L_{\rm bol}$, derived through bolometric corrections, considering an average relation between  $L_{\rm ion}$ and $L_{\rm bol}$ \citep{Markowitz2014}. However, in the present case, the simultaneous X-ray and optical/UV data from the UVOT can be used to derive a more accurate estimate of $L_{\rm ion}$. 

We considered a joint spectral analysis of the XRT and UVOT UV filters data (UVW2, UVM2, and UVW1), ignoring the optical ones that are certainly more severely contaminated by stellar light from the host galaxy. We adopted the \texttt{agnsed} model \citep{Done2012optxagnf, KubotaDone2018} for the continuum. The model assumes emission from a standard thin disc down to a radius $r_{\rm SE}$ where it is replaced by the soft X-ray excess component. Further in,  the soft-excess-emitting region is replaced by a hot corona that dominates the X-ray emission within the innermost $r_{\rm HC}$. We considered Galactic and intrinsic reddening in the UV as described by the \texttt{redden} family of \texttt{xspec} models \citep{Cardelli1989}. X-rays are absorbed using the \texttt{tbabs} and \texttt{zxipcf} models as described throughout the paper. In \texttt{agnsed}, we fixed the SMBH mass to $4.5 \times 10^{7}\ M_\odot$ \citep{Agis2014_ESO362}, and we assumed a non-spinning black hole. We also included a Gaussian emission line at $6.4$~keV as in all spectral models used so far. The system inclination was set to $i=53^\circ$.

We then applied this model to the broad-band spectra (XRT, UVW1, UVM2, and UVW2 data) during the first eclipse with the goal of deriving an average  $L_{\rm ion}$ to be used in Eq.~\ref{eq:xi}. In all cases, the statistical quality of the fits was excellent (with $C/\nu<1$). However, since the statistics is clearly driven by X-rays, we report the more interesting aspect that, in all cases, the three UV data points were well accounted for. The continuum parameters turned out to be consistent with those obtained through the \texttt{comptt} and \texttt{nthcomp} models described in the main text (see Table~\ref{tab:table4}), and we measured $kT_{\rm e} \simeq 120$~eV for the warm corona and $\Gamma \simeq 1.7$ for the hot one. Interestingly, and considering as uncertainty the spread of best-fitting values, the transition radii had median  $r_{\rm SE} = (36\pm 23)~R_{\rm g}$ and $r_{\rm HC} = (28 \pm 19)~R_{\rm g}$, broadly consistent with the ranges derived in Sects.~\ref{sec:properties} and \ref{sec:onestep}. 

\begin{figure}[t]
\centering 
\includegraphics[width=0.95\columnwidth]{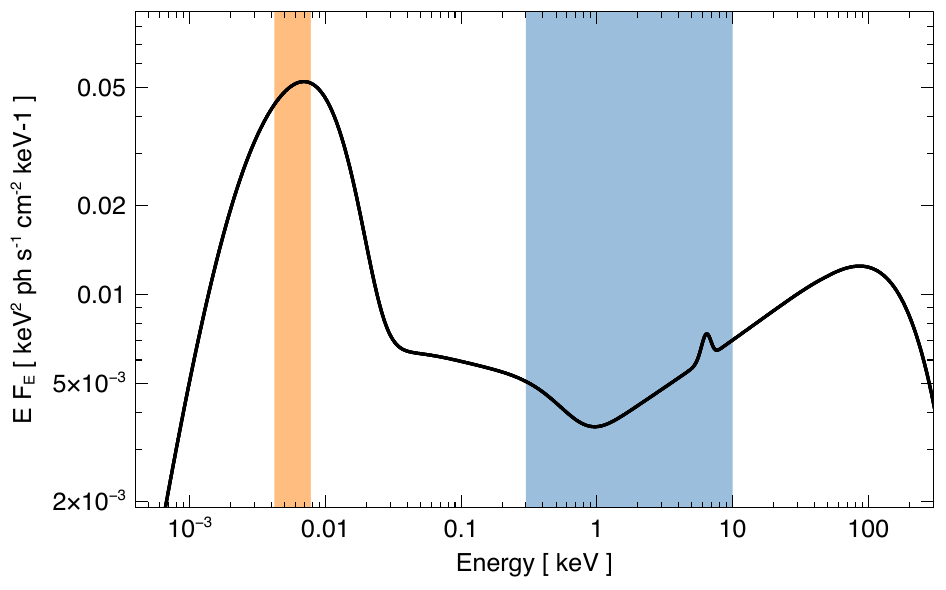}
\caption{Unabsorbed SED of ESO~362-G18 as obtained from fits using the \texttt{agnsed} model. The orange and blue shaded areas represent the bandpass of the UVOT and XRT data used in the fits. We show one example roughly corresponding to the derived averaged $L_{\rm bol}$.}
\label{fig:sed}
\end{figure}

The averaged ionising luminosity during the first eclipse was then estimated to be $L_{\rm ion} = (2.6 \pm 0.3)\times 10^{43}$~erg~s$^{-1}$. The averaged bolometric luminosity $L_{\rm bol} \simeq 1.2 \times 10^{44}\ \mathrm{erg\ s^{-1}}$, corresponds to an Eddington ratio of $\simeq 0.02$ for the assumed black hole mass, consistent with that reported by \citet{Agis2014_ESO362} based on previous X-ray observations and standard X-ray bolometric correction. In Fig.~\ref{fig:sed} we show one example of the intrinsic optical-to-X-ray spectral energy distribution (SED) of ESO~362-G18 chosen to roughly correspond to the averaged one during the first eclipse.

\subsection{A second epoch of \swift\ monitoring observations}
\label{app:X-rays_epoch2}

\begin{figure}[t]
\centering 
\includegraphics[width=0.95\columnwidth]{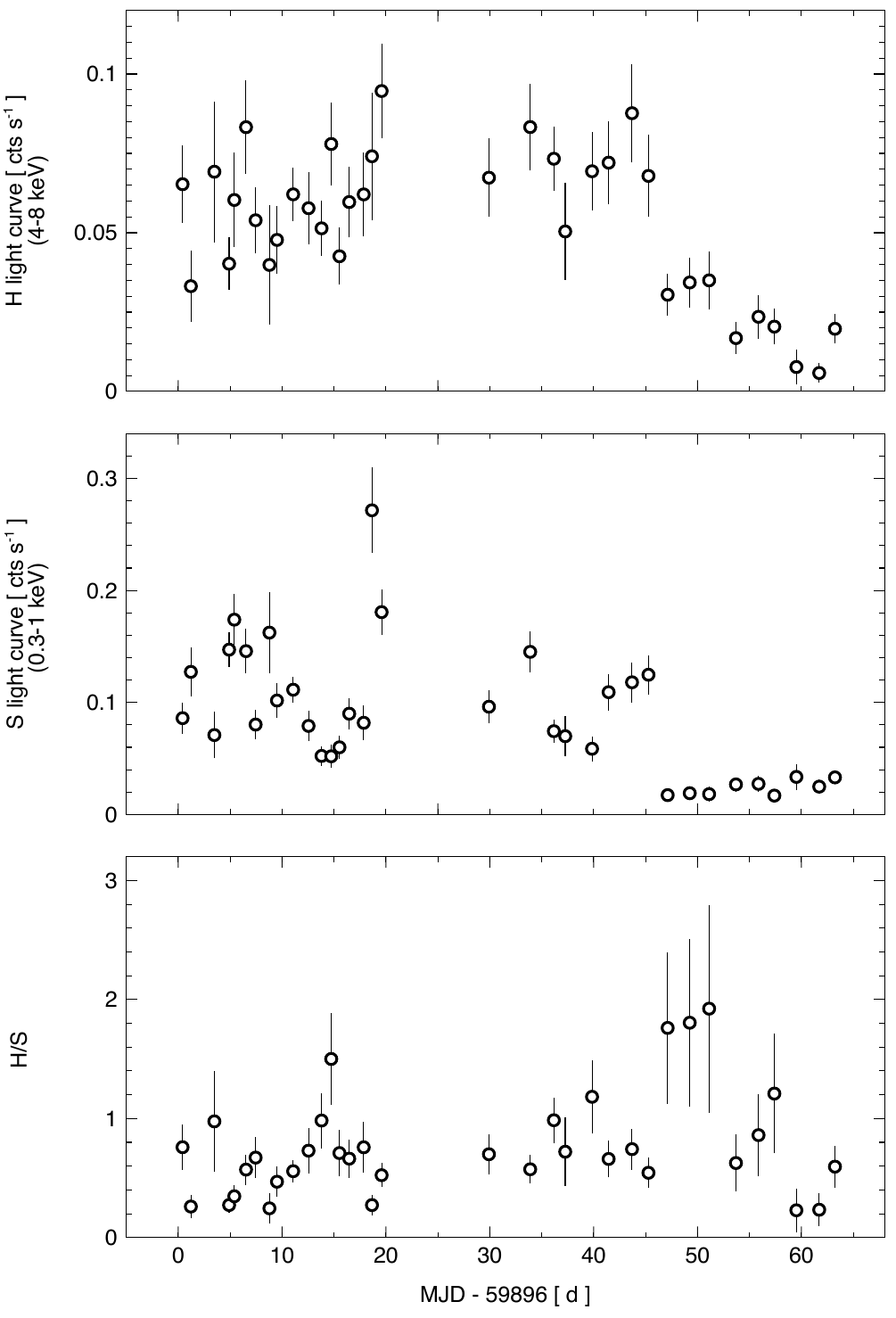}
\caption{\textit{Top and middle}: Hard (H) 4-8~keV and soft (S) 0.3-1~keV light curves during epoch~2, respectively. \textit{Bottom}: Corresponding hard-to-soft ratio (H/S).}
\label{fig:lc2}
\end{figure}

\begin{figure}
\centering 
\includegraphics[width=0.95\columnwidth]{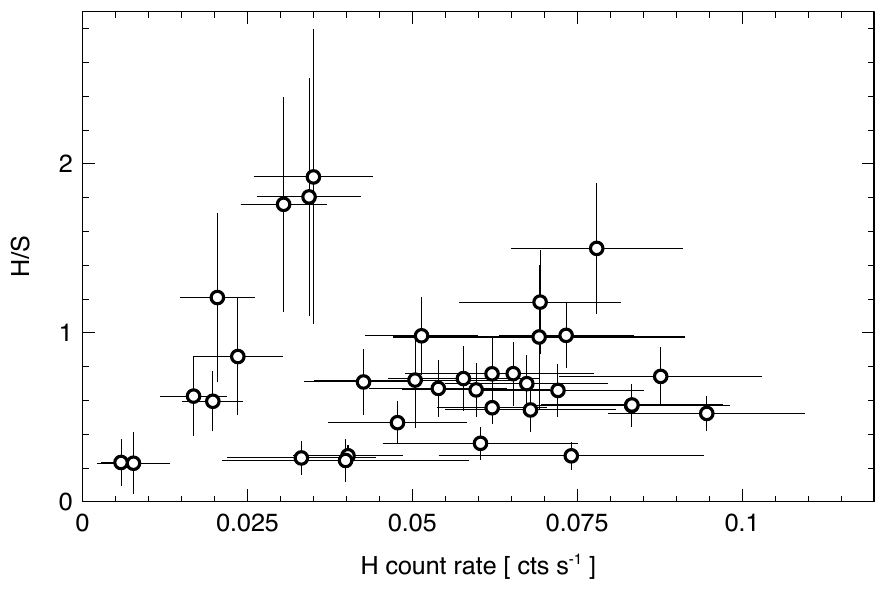}
\caption{Hard-to-soft ratio H/S as a function of hard X-ray flux during epoch~2.}
\label{fig:HS_2}
\end{figure}

ESO~362-G18 was also monitored by \swift\ during a second epoch between MJD 59896 and 59959 (hereafter epoch 2, comprising $35$ observations with a $\sim 10$~d gap in between. The H ($4$-$8$~keV), S ($0.3$-$1$~keV), and H/S light curves from epoch~2 are shown in Fig.~\ref{fig:lc2}. The hard X-ray flux during epoch~2 is about half, on average, of that during epoch~1 (see Fig.~\ref{fig:lc}). On the other hand, the H/S ratio is  generally higher by a factor of $\sim 2$ with respect to epoch~1, signalling a slightly harder spectral shape during epoch~2. The H/S ratio as a function of H count rate is shown in Fig.~\ref{fig:HS_2}. No clear correlation is seen, and the only significant excursion into a high H/S state is due to the three data points around $\simeq 50$~d in the lower panel of Fig.~\ref{fig:lc2}.

\begin{figure}[t]
\centering 
\includegraphics[width=0.95\columnwidth]{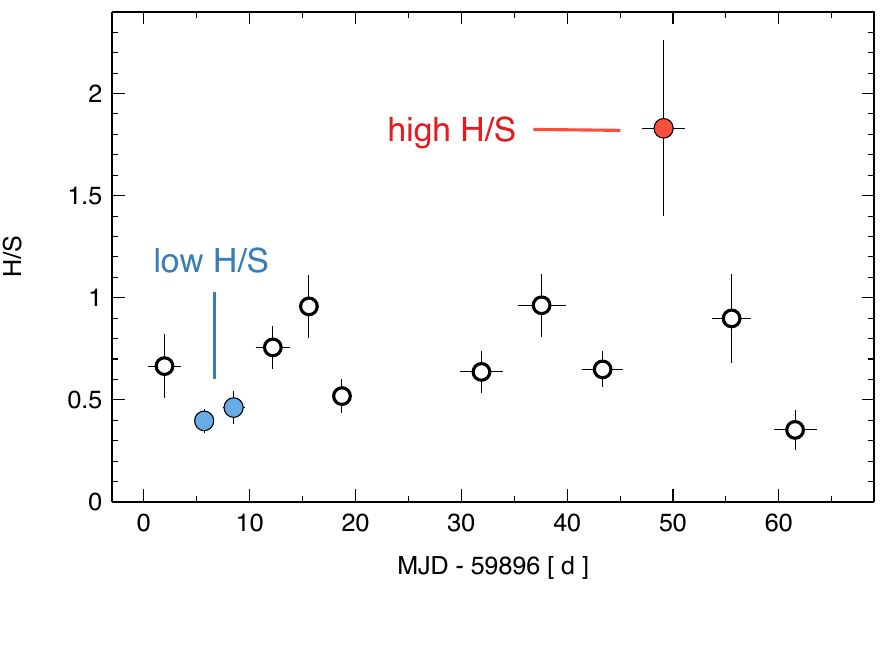}
\includegraphics[width=0.95\columnwidth]{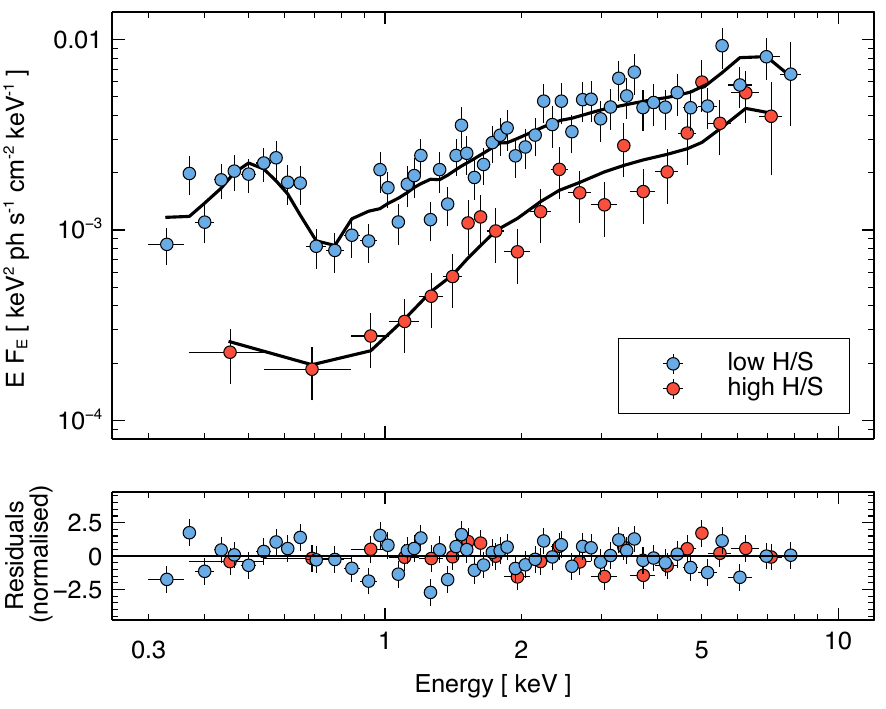}
\caption{{\it{Top:}} H/S evolution during epoch~2 re-binned by a factor of 3 except for the data point at $\sim 32$~d that is obtained by combining only two observations. Data points corresponding to particularly high (red) and low (blue) H/S are highlighted. {\it{Bottom:}} X-ray spectra extracted from the low and high H/S time intervals are shown in the upper panel together with the respective best-fitting models. The corresponding residuals are shown in the bottom panel. }
\label{fig:HS2app}
\end{figure}

The upper panel of Fig.~\ref{fig:HS2app} shows the H/S evolution during epoch~2, re-binned by a factor of 3. The H/S evolution is more erratic than during epoch~1 (compare Fig.~\ref{fig:HS2app} with Fig.~\ref{fig:epoch1_rebinned}) with only one time-interval showing a particularly high H/S. As done for epoch~1 (see Fig.~\ref{fig:HSspec}), observations corresponding to low and high H/S and from which we extracted X-ray spectra for spectral analysis, are highlighted in the figure. The resulting X-ray spectra are shown in the middle panel of Fig.~\ref{fig:HS2app} together with their respective best-fitting models and residuals, discussed below.

We first considered the low H/S spectrum and we applied the baseline model without any extra absorption, that is 
\begin{align} 
\texttt{tbabs}\times\left(\texttt{comptt}+\texttt{nthcomp}+\texttt{zgaus}\right) ~, \nonumber
\end{align}
where \texttt{tbabs} represents Galactic absorption with fixed $N_{\rm H}$. This baseline model leaves, however, significant residuals between $\sim 0.7$~keV and $\sim 1$~keV where signatures associated with warm absorbers are generally seen in low-resolution CCD X-ray spectra. We then added an ionised absorber component using the \texttt{zxipcf} model, allowing its column density, ionisation, and covering fraction to vary. After a few initial tests indicating a covering fraction $\geq 0.94$, we fixed it to $1$, that is the extra ionised absorber fully covers the X-ray source even in the low H/S spectrum (blue). The statistical improvement was $\Delta C = -78$ for two extra free parameters, and we reached a final result of $C=340$ for $404$ degrees of freedom. The absorber has $N_{\rm H} = (0.7\pm 0.3)\times 10^{22}$~cm$^{-2}$ with ionisation $\log\xi = 0.6\pm 0.9$. These parameters are typical of X-ray warm absorbers. The warm absorber explains the slightly harder spectral shape during epoch~2 with respect to epoch~1. 

We then considered simultaneous fits to the low and high HS spectra of epoch~2. As clear from Fig.~\ref{fig:HS2app}, the high H/S spectrum is more absorbed than the low H/S one, as was the case in epoch~1 (see Fig.~\ref{fig:HSspec}). We then added an extra layer of ionised absorption with the same column density and ionisation in the two spectra, but allowing the covering fraction to vary independently, that is
\begin{align} 
\texttt{tbabs}\times \texttt{zxipcf$_1$} \times \texttt{zxipcf$_2$} \times \left(\texttt{comptt}+\texttt{nthcomp} +\texttt{zgaus}\right), \nonumber
\end{align}
where \texttt{zxipcf$_1$} is a warm absorber with $C_{\rm f} = 1$ for both spectra, and \texttt{zxipcf$_2$} has the same column density and ionisation but different $C_{\rm f}$ for the low and high H/S spectra. As done for epoch~1, we assumed that any intrinsic spectral variability is negligible so that all continuum parameters were forced to be the same with an overall normalisation accounting for intrinsic flux variability. The model describes well the data with $C=467$ for 582 degrees of freedom, and the best-fitting models and residuals are shown in the middle and lower panels of Fig.~\ref{fig:HS2app}. 

The continuum parameters are fully consistent with those during epoch~1, and the extra absorber (\texttt{zxipcf$_2$}) parameters are loosely constrained to be $N_{\rm H} = (1.4\pm 1.1)\times 10^{22}$~cm$^{-2}$ and $\log\xi = (-0.2\pm 1.0)$. The covering fraction is $C_{\rm f} \leq 0.4$ in the low H/S spectrum, and $C_{\rm f} \geq 0.6$ in the high H/S one. The small contrast in covering fraction between the low and high H/S spectra prevented us from performing a detailed analysis as done during the first eclipse in epoch~1. Moreover, the H/S evolution during epoch~2 is much more erratic than during epoch~1 and appears to be characterized by significant fluctuations. This  might suggest fast variability of the warm absorber rather than a series of very short-duration eclipses by individual clouds. 

\section{UVOT data analysis}
\label{app:UVOT}

\begin{figure*}[t]
\centering 
\includegraphics[width=0.9\columnwidth]{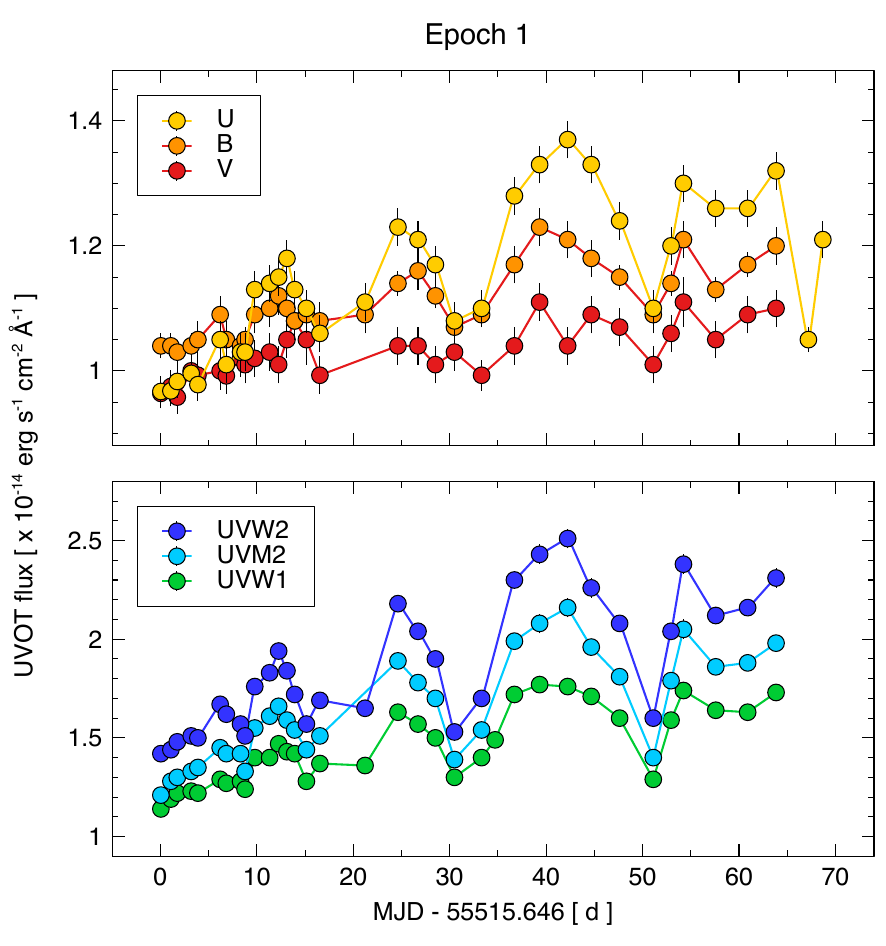}
{\hspace{0.1cm}}
\includegraphics[width=0.9\columnwidth]{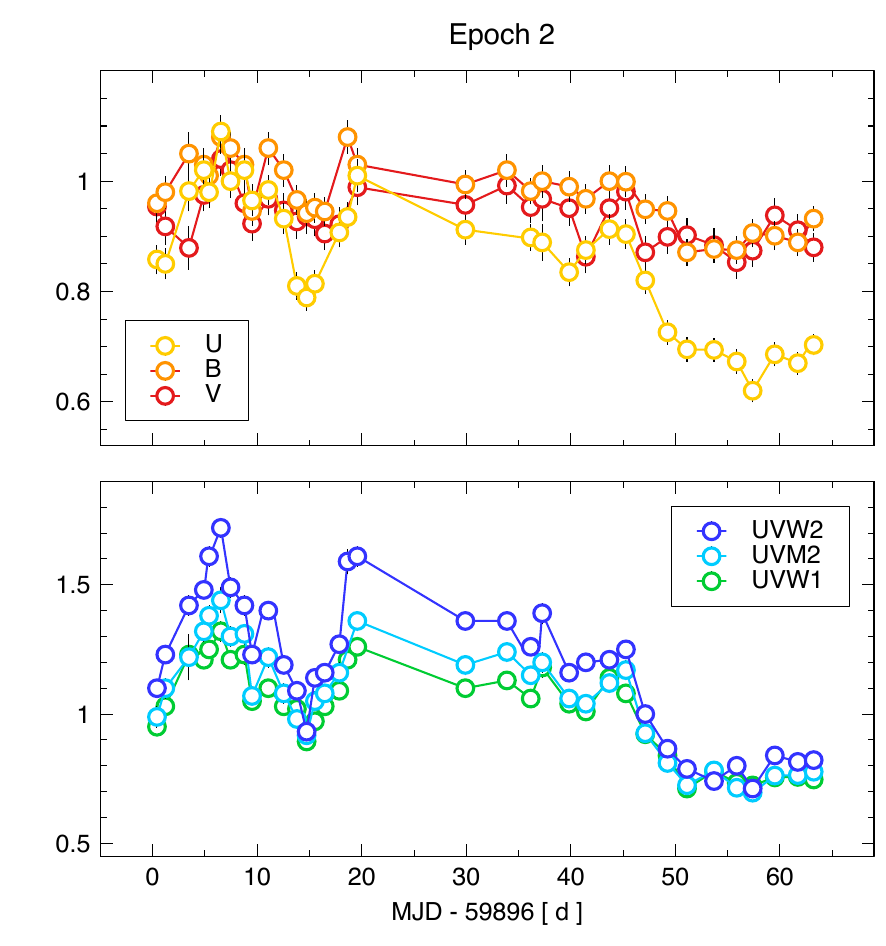}
\caption{UVOT light curves in all filters for epoch~1 (\textit{left}) and epoch~2 (\textit{right}). Optical (V, B, and U) and UV (UVW1, UVM2, UVW2) filters have been separated for better visual clarity. Flux errors for the UV filters (lower panels) are smaller than the symbol size.}
\label{fig:UVOT_lc}
\end{figure*}

\begin{figure*}[t]
\centering 
\includegraphics[width=0.9\columnwidth]{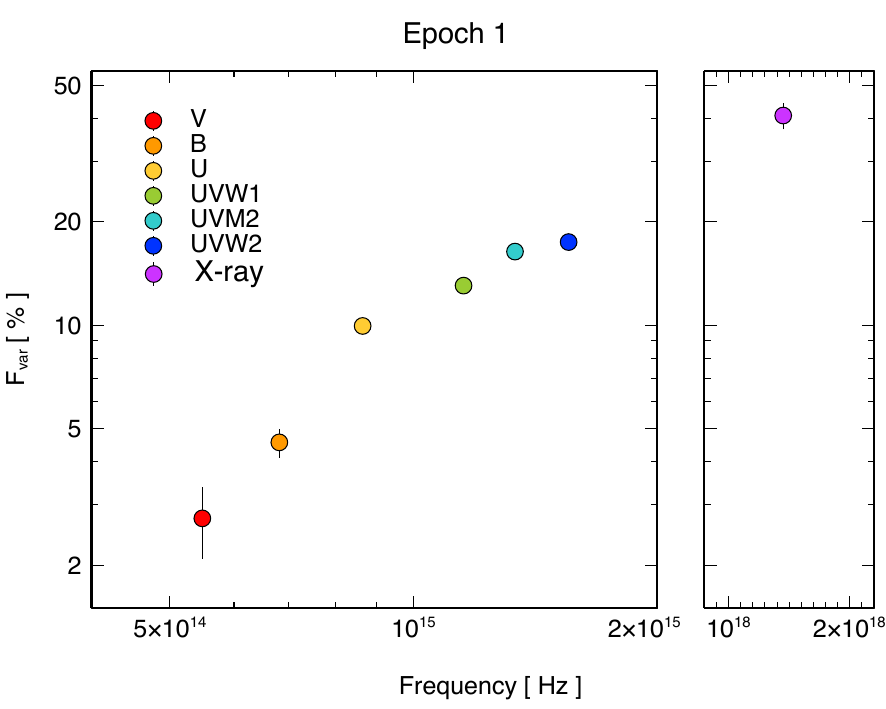}
{\hspace{0.2cm}}
\includegraphics[width=0.9\columnwidth]{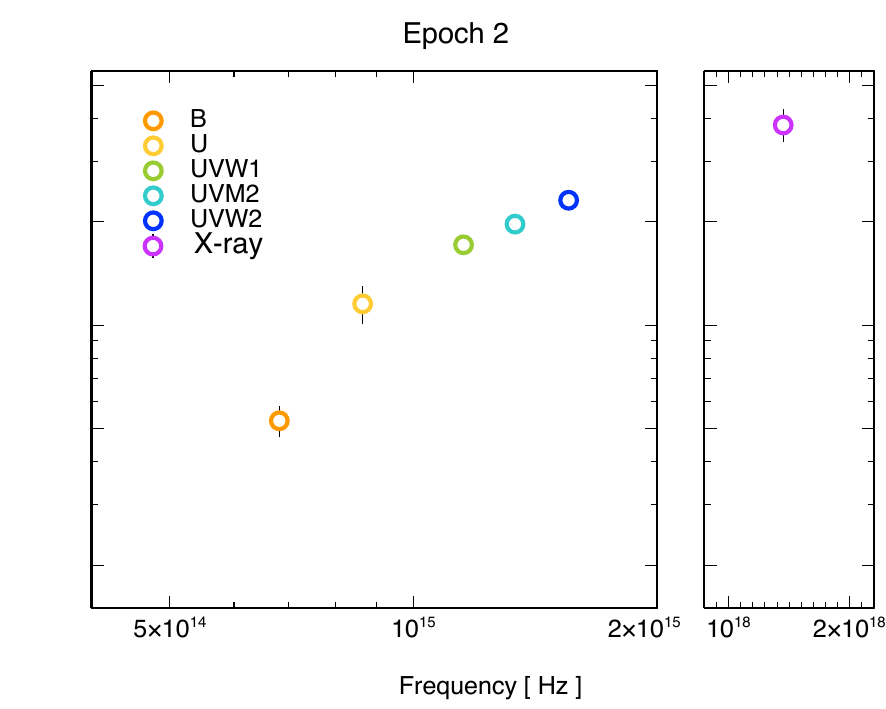}
\caption{Fractional variability amplitude $F_{\rm var}$ as a function of frequency for epoch~1 (\textit{left}) and epoch~2 (\textit{right}). The reddest V filter for epoch~2 has $F_{\rm var} \leq 1.1$\% and is not shown here for visual clarity. The y-axis scale is the same in all panels.}
\label{fig:Fvar}
\end{figure*}

Nuclear optical and UV fluxes from UVOT exposures simultaneous with the XRT ones were extracted from the corresponding images in all six filters (V, B, U, UVW1, UVM2, UVW2) as described in Sect.~\ref{sec:obs}. The UVOT light curves for epochs~1 and 2 are shown in Fig~\ref{fig:UVOT_lc}. During epoch~2, the UVOT fluxes are all about a factor of $2$ lower than during epoch~1, as was the case for hard X-rays, most likely signalling a lower mass accretion rate. The UVOT light curves are variable, and Fig.~\ref{fig:Fvar} shows the fractional variability amplitude $F_{\rm var}$ - as defined by  \citet{Vaughan2003} - as a function of frequency, showing a trend of increasing variability with frequency on the probed timescales. We point out that part of the trend is certainly due to increasing dilution by stellar light towards the longest wavelengths since the adopted UVOT aperture of 5\arcsec corresponds to $\sim 1.3$~kpc at the redshift of ESO~362-G18. 

\begin{figure*}[t]
\centering 
\includegraphics[width=1.8\columnwidth]{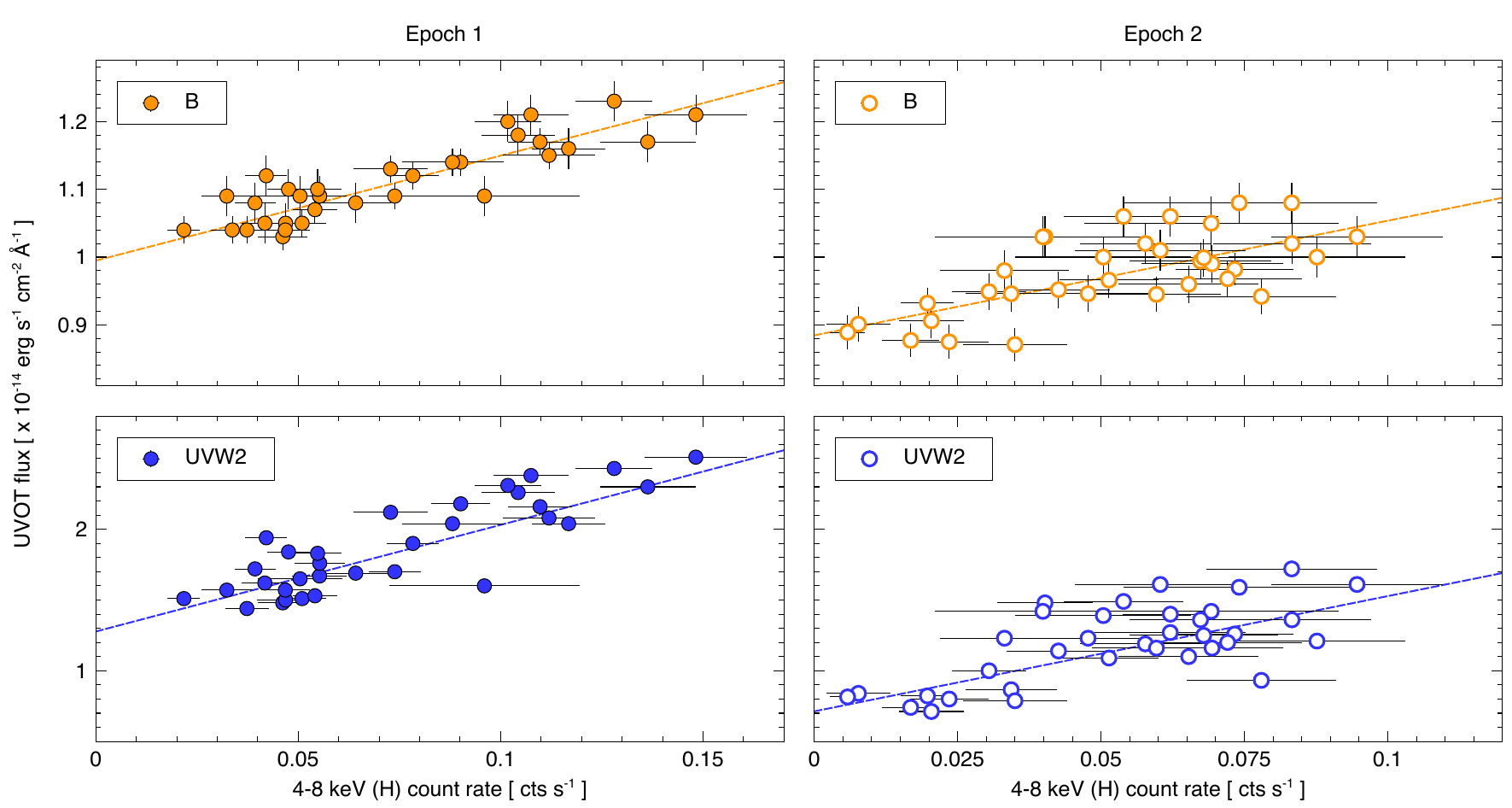}
\caption{B and UVW2 fluxes as a function of hard X-ray count rate (H) for both epochs. The dashed lines show the best-fitting linear relation. The y-axis for each filter is the same. The flux errors in the lower panels are smaller than the symbol size. 
}
\label{fig:X-uvot-corr}
\end{figure*}

A visual comparison of  Fig.~\ref{fig:lc} and ~\ref{fig:UVOT_lc} indicates that the hard X-ray and UVOT light curves are relatively well correlated. This is shown quantitatively in Fig.~\ref{fig:X-uvot-corr} for both epochs and for one optical (B) and one UV (UVW2) filter. Pearson correlation coefficients (r) for the three UV filters (UVW1, UVM2, and UVW2) and for both epochs are in the range $r_{\rm UV}\simeq 0.81$-$0.86$. The reddest V-filter show the least correlation with $r_{\rm V} \simeq 0.5$-$0.6$, while $r_{\rm U} = 0.75$-$0.83$, and $r_{\rm B} \simeq 0.82$-$0.86$. We have attempted to compute time lags between the hard X-rays and UVOT light curves using the \texttt{Pycorrelate} code\footnote{\url{https://pycorrelate.readthedocs.io/en/latest/index.html}}, but all lags were found to be consistent with zero at both epochs (as well as considering combined time series), likely because of insufficient cadence. 

In both epochs, the UVOT light curves are characterised by dips, typically lasting a few days, that are more clearly seen in the UV than in the optical filters, although the bluest U filter sometimes shows similar behaviour to the UV ones. The two best-defined flux drops are seen during epoch~2 around $\sim 15$~d and $\gtrsim 45$~d into the campaign; see the right panels of Fig.~\ref{fig:UVOT_lc}. Given the detection of X-ray eclipses in ESO~362-G18, and having established that these are likely due to clouds close to the dust sublimation zone where gas and dust co-exist, it is tempting to associate the UV dimming with transient occultation events \citep[see also][]{Agis2014_ESO362}. However, the lack of optical and UV spectra, which could be used to disentangle intrinsic optical/UV variability from extrinsic events, prevented us from performing a detailed analysis of these events as done in the X-rays.

\end{appendix}

\end{document}